\documentclass{article}
\usepackage[a4paper, total={6.3in, 9in}]{geometry}
\usepackage{graphicx,fancyhdr,tabularx,amsmath,amssymb,color, bm}
\usepackage{subcaption}
\usepackage{physics}
\usepackage{breqn}
\usepackage[export]{adjustbox}
\usepackage{url}
\usepackage{float}
\usepackage{svg}
\usepackage{titlesec, blindtext, color}
\definecolor{gray75}{gray}{0.75}
\usepackage{verbatim}
\usepackage{multirow}
\usepackage{titlesec}
\usepackage{todonotes}
\usepackage{sidecap} 

\NewDocumentCommand{\evalat}{sO{\big}mm}{%
  \IfBooleanTF{#1}
   {\mleft. #3 \mright|_{#4}}
   {#3#2|_{#4}}%
}

\usepackage[
backend=biber,
date=year,
sorting=none,
defernumbers=true,
giveninits=true,
maxbibnames=10,
doi=false,
isbn=false,
url=false,
eprint=false,
arxiv=false,
natbib=true]{biblatex}
\AtEveryBibitem{%
  \clearfield{note}%
  \clearfield{volume}
  \clearfield{number}
  \clearfield{series}
}
\DeclareSourcemap{ 
  \maps{
    \map{
      \step[fieldsource=language, fieldset=langid, origfieldval, final]
      \step[fieldset=language, null]
    }
  }
}
\makeatletter
\let\blx@rerun@biber\relax
\makeatother

\usepackage{listings}
\usepackage{xcolor}

\usepackage{authblk} 

\definecolor{codegreen}{rgb}{0.2,0.7,0.2}
\definecolor{codeblue}{rgb}{0,0.2,0.5}
\definecolor{codered}{rgb}{0.7,0.3,0.3}
\definecolor{backcolour}{rgb}{0.95,0.95,0.92}
\lstdefinestyle{mystyle}{
    backgroundcolor=\color{backcolour},   
    commentstyle=\color{codeblue},
    keywordstyle=\color{codegreen},
    stringstyle=\color{codered},
    basicstyle=\ttfamily\footnotesize,
    breakatwhitespace=false,         
    breaklines=true,                 
    captionpos=b,                    
    keepspaces=true,                 
    numbers=left,                    
    numbersep=5pt,                  
    showspaces=false,                
    showstringspaces=false,
    showtabs=false,                  
    tabsize=2
}
\newcommand{\vv}[1]{\boldsymbol{#1}}

\begin{document}
\title{Pulsatory patterns in active viscoelastic fluids with distinct relaxation time scales}
\author[1,2]{E. M. de Kinkelder}
\author[3,4,5]{E. Fischer-Friedrich \thanks{elisabeth.fischer-friedrich@tu-dresden.de}}
\author[1,2]{S. Aland \thanks{sebastian.aland@math.tu-freiberg.de}}
\affil[1]{Institute of Numerical Analysis and Optimization, Technische Universtit\"at Bergakademie Freiberg, Freiberg, Germany}
\affil[2]{Faculty of Informatics/Mathematics, Hochschule f\"ur Technik und Wirtschaft, Dresden, Germany}
\affil[3]{Cluster of Excellence Physics of Life, Technische Universit\"at Dresden, Dresden, Germany}
\affil[4]{Biotechnology Center, Technische Universit\"at Dresden, Dresden, Germany}
\affil[5]{Faculty of Physics, Technische Universit\"at Dresden, Dresden, Germany}

\maketitle
\begin{abstract}
Developing tissues need to pattern themselves in space and time. A prevalent mechanism to achieve this are pulsatile active stresses generated by the actin cytoskeleton. Active gel theory is a powerful tool to model the dynamics of cytoskeletal pattern formation. In theoretical models, the influence of the viscoelastic nature of the actin cytoskeleton has so far only been investigated by the incorporation of one viscoelastic relaxation time scale.  Here, using a minimal model of active gel theory  with a single molecular regulator, we show that distinct shear and areal relaxation times are sufficient to drive pulsatile dynamics in active surfaces.
\end{abstract}

\section{Introduction}
Pattern formation and self-deformation of active gels have become increasingly recognised to be an essential contribution to our understanding of the dynamics and morphogenesis of living systems \cite{prost2015, joanny2009}. In mammalian cells and tissues, pattern formation and force generation often rely on the actin cytoskeleton. This intracellular biopolymer network constitutes an active gel that can generate active contractile stresses through the activity of ATP-consuming molecular motor proteins that bind to it. In this way, a sheet-like actin cytoskeleton can convey an actively regulated surface tension to cellular interfaces that, in turn, may drive cell shape changes. Accordingly, contractile stress and advective fluxes in the actin cytoskeleton are known to be vital for the processes of mitotic rounding and cytokinesis during cell division \cite{taubenberger2020, wagn16,salb12, wittwer2022computational}. 

Experimental research observed a variety of oscillatory actin cytoskeletal dynamics in cells and tissues \cite{Bailles2019, Allard2013, mitsushima2010, Wu2013}.
In particular, previous studies showed that pulsatile oscillatory time dynamics is an important feature of many morphogenetic processes \cite{gorfinkiel2016, blanchard2018, miao2020}.
Theoretical models could reproduce pulsatile patterns in active gels by i) incorporating at least two molecular species and nonlinearities in their mechanochemical regulation \cite{Staddon2022, kuma14, bonati2022}, or by ii) combining active hydrodynamics with the dynamics of a polarization vector field \cite{Liu2021, Marcq2014}, by iii) coupling an axisymmetric viscoelastic active surface with a surrounding highly viscous fluid \cite{MietkeThesis2018}. 

So far, many studies on active gel theory have been focusing on viscous active surfaces \cite{mietke2019, mietke2019b, bois11, kuma14, salb17, salb09, bert14, gros19,reym16, prost2015}. However, experimental measurements have shown that the actin cytoskeleton is a viscoelastic material, whose resistance to shear deformation becomes fluid-like on time scales beyond minutes, likely due to molecular turnover \cite{bonfanti2020, hoss21, salb12}. With regards to areal deformations, viscoelastic relaxation times may be influenced by active cell surface area regulation, e.g. through exocytosis and endocytosis. Such active surface area regulation may convey areal surface elasticity on significantly longer time scales. Correspondingly, viscoelastic relaxation times of shear and bulk mechanics in the actin-cytoskeletal films may be distinct from each other. To date, the influence of such distinct time scales on the dynamics of active surfaces remains however elusive.

Here, using FEM simulations and linear stability analysis, we study flat active viscoelastic surfaces with an active component. We model the viscoelasticity with the upper convected surface Maxwell model using  distinct relaxation time scales for shear and areal surface mechanics \cite{DeKinkelder2021}. We show that different relaxation times for the shear and areal stress by themselves may lead to complex pulsatile pattern formation in the system, even if only a single molecular regulator is present.

\section{Governing equations}
We consider a 2-dimensional surface $\Omega = [0,L]^2$ with periodic boundary conditions. The surface is viscoelastic and carries a surface bound species, which induces an active stress. The viscoelastic stress is modelled by the upper convected Maxwell model on surfaces as described in \cite{DeKinkelder2021}. The model distinguishes between shear and bulk stress. Accordingly, the stress tensor $S$ is split in the bulk stress represented by $\tr (S)$ and the shear stress represented by the traceless part $\bar{S} = S - \frac{\tr (S)}{2} I$, where $I$ is the identity matrix. The equations for the stress are
\begin{align}
    \bar{S} &= 2 \eta_S \bar{D} - \tau_S \left(\partial_{t}^\bullet \bar{S} - \nabla \vv{v} \bar{S} - \bar{S} (\nabla \vv{v})^T + I(\bar{S} : \nabla \vv{v}) - \tr (S) \bar{D} \right), \label{eq:devS}\\
    \tr(S) &= 2 \eta_B \tr(D) + \tau_B \left(2(\bar{S}:\nabla \vv{v}) + \tr (S)\tr (D) -\partial_{t}^\bullet \tr (S) \right), \label{eq:trS}
\end{align}
where $\vv{v}$ is the velocity and $D$ is the rate of deformation, $D = \frac{1}{2} \left( \nabla \vv{v} + (\nabla \vv{v})^T \right)$. Similarly to the stress tensor, the traceless rate of deformation $\bar{D}$ is defined as $\bar{D} = D - \frac{\tr (D)}{2} I$. $\partial_t^\bullet$ is the material derivative. The parameters are the bulk and shear viscosity $\eta_B$, $\eta_S$ and the bulk and shear relaxation times $\tau_B$, $\tau_S$, which give rise to the corresponding elastic moduli $G_B = \frac{\eta_B}{\tau_B}$, $G_S = \frac{\eta_S}{\tau_S}$. Further, we assume a single regulator which is described by concentration $c$. The dynamics of the regulator are defined by a convection-diffusion equation,
\begin{equation}
    \partial_t c + \nabla \cdot (c \vv{v}) = D_c \Delta c,
    \label{eq:concentration}
\end{equation}
where $D_c$ is the diffusion constant. The presence of the regulator initiates an isotropic contractile stress which we model by $\xi f(c)$. Here, $\xi$ scales the activity and $f(c) = \frac{c^2}{c^2 + c_0^2}$, where $c_0$ is the equilibrium concentration. Accordingly, the equations for the stress (Eqs.~\eqref{eq:devS}-\eqref{eq:trS}) and the concentration (Eq.~\eqref{eq:concentration}) are coupled with the following force balance,
\begin{equation}
    \rho  \partial_t^\bullet \vv{v} = \nabla \cdot (S + \xi f(c) I), \label{eq:forceBalanceFull}
\end{equation}
where $\rho$ is the mass density. To scale the equations, we use the width of the domain $L$ as length scale and the diffusive time scale ${L^2}/{D_c}$. Additionally, we define the following dimensionless parameters in the viscoelastic equations $\hat{G}_{\alpha} = {G_{\alpha}}/{\xi}$, $\hat{\tau}_{\alpha} = {\tau_{\alpha} D_c}/{L^2}$ for $\alpha = B,S$. These then define the scaled viscosities $\hat{\eta}_{\alpha} = \hat{G}_{\alpha}\hat{\tau}_{\alpha}$. In the force balance, we get the scaled density $\hat{\rho} = {D_c^2 \rho}/({L^2 \xi})$. The dimensionless equations are
\begin{align}
    \frac{1}{\hat{\tau}_S} \bar{S} &= 2 \hat{G}_S \bar{D} - \left(\partial_{t}^\bullet \bar{S} - \nabla \vv{v} \bar{S} - \bar{S} (\nabla \vv{v})^T + I(\bar{S} : \nabla \vv{v}) - \tr (S) \bar{D} \right), \label{eq:devS_scaled}\\
    \frac{1}{\hat{\tau}_B} \tr(S) &= 2 \hat{G}_B \tr(D) + \left(2(\bar{S}:\nabla \vv{v}) + \tr (S)\tr (D) -\partial_{t}^\bullet \tr (S) \right), \label{eq:trS_scaled}\\
    \partial_t c + \nabla \cdot (c \vv{v}) &= \Delta c, \label{eq:c_scaled}\\
    \hat{\rho} \left( \partial_t^\bullet \vv{v} \right) &= \nabla \cdot (S + f(c) I), \label{eq:force_scaled}
\end{align}
For readability, we will omit the hat on the scaled parameters for the rest of the article.
Note, that in the limit of large relaxation times ($\tau_B, \tau_S \rightarrow \infty$) the model corresponds to Neo-Hookean surface elasticity, while for small relaxation times ($\tau_B, \tau_S \rightarrow 0$, $\eta_B, \eta_S\in \mathcal{O}(1)$) and small $\hat{\rho}$ it approaches the compressible Navier-Stokes momentum equation.

\section{Linear stability analysis}
To study pattern formation, we perform a linear stability analysis. For this, we use the linearised version of the viscoelastic stress terms in Eqs.~\eqref{eq:devS_scaled} and \eqref{eq:trS_scaled}
\begin{align}
    \frac{1}{\tau_S}\bar{S} &= 2 G_S \bar{D} - \partial_t \bar{S}, \label{eq:devSlin}\\
    \frac{1}{\tau_B}\tr(S) &= 2 G_B \tr(D) - \partial_{t}\tr (S). \label{eq:trSlin}
\end{align}
As we model the regime of low Reynolds numbers \cite{purcell1977}, the density $\rho$ in Eq.~\eqref{eq:force_scaled} is assumed to be very small, so the force balance reduces to $\nabla \cdot (S + \xi f(c) I) = 0$. For the linear stability analysis, we assume the variables to be equal to a stationary value plus a small perturbation such that $c = 1 + \delta c$, $\vv{v} = \vv{0} + \delta \vv{v}$, $\tr (S) = 0 + \delta \tr (S)$ and $\bar{S} = 0 + \delta \bar{S}$. Here, the fields $\delta c$, $\delta \vv{v}$, $\delta \tr (S)$ and $\delta \bar{S}$ are considered as  small perturbations defined as
\begin{align}
    \delta c &= \sum_{\vv{k}} \delta c^{\vv{k}} \exp(i \vv{k} \cdot \vv{x}) \exp(\lambda_{\vv{k}} t), \label{eq:c_perturb} \\
    \delta \vv{v} &= \sum_{\vv{k}} \delta \vv{v}^{\vv{k}} \exp(i \vv{k} \cdot \vv{x}) \exp(\lambda_{\vv{k}} t), \label{eq:v_perturb} \\
    \delta \tr (S) &= \sum_{\vv{k}} \delta \tr (S)^{\vv{k}} \exp(i \vv{k} \cdot \vv{x}) \exp(\lambda_{\vv{k}} t), \label{eq:trS_perturb} \\
    \delta \bar{S} &= \sum_{\vv{k}} \delta \bar{S}^{\vv{k}} \exp(i \vv{k} \cdot \vv{x}) \exp(\lambda_{\vv{k}} t), \label{eq:Sbar_perturb} 
\end{align}
where the wave vector $\vv{k} = \frac{2 \pi}{L} \vv{n}$ with $\vv{n} \in \mathbb{N}\times \mathbb{N}$. The coefficients $\delta c^{\vv{k}}, \delta \tr (S)^{\vv{k}} \in \mathbb{C}$, $\delta \vv{v}^{\vv{k}}\in \mathbb{C}^2$ and $\delta \bar{S}^{\vv{k}} \in \mathbb{C}^{2\times 2}$ are constant and $\lambda_{\vv{k}} \in \mathbb{C}$ is the growth rate for the respective mode. For readability, we will omit the subscript $\vv{k}$ in the eigenvalue $\lambda_{\vv{k}}$. Substituting the perturbation ansatz for the concentration and velocity field into Eq.~\eqref{eq:c_scaled} and considering the term for each mode  separately, we obtain
\begin{equation}
    \lambda \delta c^{\vv{k}} + i \vv{k} \cdot \delta \vv{v}^{\vv{k}} = - \vv{k}^2 \delta c^{\vv{k}}.
    \label{eq:concentrationPerturb}
\end{equation}
Further, substituting Eqs.~\eqref{eq:v_perturb}-\eqref{eq:Sbar_perturb} into the linear viscoelastic stress equations (Eqs.~\eqref{eq:devSlin} and \eqref{eq:trSlin}) results in the following two equations,
\begin{align}
    \delta \tr (S)^{\vv{k}} &= \frac{2iG_B}{\lambda + 1/\tau_B} \vv{k} \cdot \delta \vv{v}^{\vv{k}}, \label{eq:trS_eq_perturb}\\
    \delta \bar{S}^{\vv{k}} &= \frac{iG_S}{\lambda + 1/\tau_S} \left( (\delta \vv{v}^{\vv{k}})^T\vv{k} + \vv{k}^T \delta \vv{v}^{\vv{k}} - \vv{k} \cdot \delta \vv{v}^{\vv{k}} I \right). \label{eq:Sbar_eq_perturb}
\end{align}
Using $S = \bar{S} + \frac{\tr (S)}{2} I$, we can substitute Eqs.~\eqref{eq:trS_eq_perturb} and \eqref{eq:Sbar_eq_perturb} into the linear force balance. This allows us to derive an expression that does not depend on the perturbations of the viscoelastic stress,
\begin{equation}
    0 = -\frac{G_S}{\lambda + 1/\tau_S} \vv{k}^2 \delta \vv{v}^{\vv{k}} - \frac{G_B}{\lambda + 1/\tau_B} \vv{k} \cdot \delta\vv{v}^{\vv{k}} \vv{k} + i f'(1) \delta c^{\vv{k}} \vv{k}.\label{eq:v_dot_kVE1}
\end{equation}
Then, by taking the inner product with $\vv{k}$ and dividing by $\vv{k}^2$, we derive an expression for $\delta \vv{v}^{\vv{k}} \cdot \vv{k}$,
\begin{equation}
   \delta \vv{v}^{\vv{k}} \cdot \vv{k} = \frac{\left(\lambda + \frac{1}{\tau_B}\right)\left(\lambda + \frac{1}{\tau_S}\right)}{G_S\left(\lambda + \frac{1}{\tau_B}\right) + G_B \left(\lambda + \frac{1}{\tau_S}\right)} i f'(1) \delta c^{\vv{k}}.
    \label{eq:v_dot_kVE2}
\end{equation}
When we substitute this into Eq.~\eqref{eq:concentrationPerturb}, the eigenvalue $\lambda$ can be calculated. Before doing so however, we first distinguish two cases: i) equal relaxation times $\tau_B = \tau_S$ and ii) unequal relaxation times $\tau_B \neq \tau_S$. For the first case, the eigenvalues are
\begin{equation}
    \lambda = \frac{f'(1)/\tau - \vv{k}^2}{G_S + G_B - f'(1)}.
\end{equation}
The derivation can be found in App.~\ref{app:equalRelaxationTimes}. With this definition, the eigenvalue $\lambda$ cannot be complex. Hence the solution does not oscillate.
For the second case with $\tau_B \neq \tau_S$, substituting Eq.~\eqref{eq:v_dot_kVE2} into Eq.~\eqref{eq:concentrationPerturb} results in a quadratic equation for the eigenvalues. We will write this as $a \lambda^2 + b\lambda + c = 0$, with the coefficients $a$, $b$, $c$ defined as
\begin{align}
    a &= \tau_B \tau_S (G_S + G_B - f'(1)), \label{eq:a}\\
    b &= \tau_B G_B + \tau_S G_S + \tau_B \tau_S (G_B+G_S) \vv{k}^2 -(\tau_B+\tau_S)f'(1), \label{eq:b} \\
    c &= (\tau_B G_B + \tau_S G_S) \vv{k}^2 -f'(1). \label{eq:c}
\end{align}
The eigenvalues are the zeros of this quadratic equation. This potentially results in two eigenvalues $\lambda^-$ and  $\lambda^+$ defined as  
\begin{dmath}
\scalebox{1.1}{$\lambda^\pm = -\frac{\tau_B G_B + \tau_S G_S + \tau_B \tau_S (G_B+G_S) \vv{k}^2 -(\tau_B+\tau_S)f'(1)}{2\tau_B \tau_S (G_S + G_B - f'(1))} \pm$}\\
\scalebox{1.1}{$ \frac{\sqrt{\left( \tau_B G_B + \tau_S G_S + \tau_B \tau_S (G_B+G_S) \vv{k}^2 -(\tau_B+\tau_S)f'(1) \right)^2 - 4\tau_B \tau_S (G_S + G_B - f'(1))\left( (\tau_B G_B + \tau_S G_S) \vv{k}^2 -f'(1) \right)}}{2\tau_B \tau_S (G_S + G_B - f'(1))}.$}
\label{eq:lambda}
\end{dmath}
For both equal and unequal relaxation times, the eigenvalues can become arbitrarily large if $G_B + G_S < f'(1)$. In the full system of equations Eqs.~\eqref{eq:devS_scaled}-\eqref{eq:force_scaled}, these small wavelengths are suppressed by the nonlinear terms. For the linear stability analysis however, the following condition is required to prevent arbitrarily large eigenvalues
\begin{equation}
    G_B + G_S > f'(1).
    \label{eq:mainCondition}
\end{equation}
So for the remainder of the linear stability analysis, we assume that $G_B + G_S > f'(1)$. This can be interpreted as the surface elasticity has to be stronger than the active surface tension to allow oscillations. 

For the second case with unequal relaxation times, oscillations only occur if the discriminant $\mathcal{D}$ of the quadratic equation is negative (the derivation is found in App.~\ref{app:complexEig}). We found that this can only be if the condition in Eq.~\eqref{eq:mainCondition} holds.
Note that Eq.~\eqref{eq:mainCondition} does not guarantee that there will be oscillations. It only implies that there exists an interval, $[k^-, k^+]$ with $k^-, k^+ \in \mathbb{R}$ and $\mathcal{D}_{|\vv{k}^2 = k^-} = \mathcal{D}_{|\vv{k}^2 = k^+} = 0$, such that  $\lambda_{\vv{k}}$ is only complex for wave vectors $\vv{k}$ with $\vv{k}^2 \in [k^-, k^+]$.

\subsection{Stability}
\label{sec:stability}
For the stability of the system, we are interested in the sign of the real part of the largest eigenvalue, i.e. $\text{Re}(\lambda^+) = \text{Re}\left( \frac{-b + \sqrt{\mathcal{D}}}{2a} \right)$. We show that the largest eigenvalue is always assumed for the excitable mode with the lowest value of $\vv{k}^2$. For $\vv{k}^2=0$, we have to have $\delta c^{\vv{0}} = 0$  due to particle conservation  (see App.~\ref{app:0mode}). Therefore, the lowest mode corresponds to $\vv{n}^2 = 1$ ($\vv{k}^2 = 4 \pi^2$), which is assumed for either $\vv{n} = \begin{pmatrix} 1\\0 \end{pmatrix}$ or for $\vv{n} = \begin{pmatrix} 0\\1 \end{pmatrix}$. We will refer to either of these modes as 1-mode.
To calculate the stability of the system, only the eigenvalue of the 1-mode $\lambda^+_{|\vv{n}^2 =1}$ needs to be considered. The condition $G_S + G_B > f'(1)$ implies that $a>0$. The real part of the eigenvalue $\text{Re}(\lambda^+) = \text{Re}(\frac{-b + \sqrt{b^2-4ac}}{2a})$ is greater than 0 if $b<0$ or $c<0$. This results in two conditions. The first condition, $b<0$, implies that
\begin{equation}
f'(1) > \frac{\tau_B G_B + \tau_S G_S + 4 \pi^2 \tau_S \tau_B (G_B + G_S)}{\tau_B + \tau_S}
\label{eq:b<0}
\end{equation}
Using the relation $\eta_{B/S} = G_{B/S} \tau_{B/S}$, we can rewrite Eq.~\eqref{eq:b<0} to 
\begin{equation}
f'(1) > \frac{\tau_B}{\tau_B + \tau_S}\left(G_B + 4 \pi^2 \eta_S \right) + \frac{\tau_S}{\tau_B + \tau_S}\left(G_S + 4 \pi^2 \eta_B \right).
\label{eq:b>0 2}
\end{equation}
This can be interpreted as the system is unstable if the active surface tension is stronger than a weighted average of the viscous and elastic stress. The weight is defined by the relaxation times $\tau_B$ and $\tau_S$. So if $\tau_B \gg \tau_S$, meaning that we have a bulk elastic component and a shear viscous component, then the stability mainly depends on both of these components. The second condition, $c<0$, implies that
\begin{equation}
f'(1) > 4 \pi^2 (G_B \tau_B + G_S \tau_S) = 4 \pi^2 (\eta_B + \eta_S).
\label{eq:c>0}
\end{equation}
This can be interpreted as the system is unstable if the active surface tension is stronger than the viscosity. When this condition is met, however, there can be no complex eigenvalues, i.e. no oscillations. Both conditions also imply that a purely elastic surface will result in a stable solution. Additionally, there are two more important conclusions to draw from the linear stability analysis. The first is that the shear and bulk components are interchangeable. Looking at the eigenvalues in Eqs.~\eqref{eq:lambda} and \eqref{eq:lambdaEta}, we see that replacing all bulk parameters by shear parameters and vice versa results in the same eigenvalues. The second conclusion is that making another choice for the function $f$ will give the same results in the linear stability analysis, as long as $f'(1)>0$.

\section{Dynamics in the linear regime}
\begin{figure}
    \centering
    \begin{subfigure}[t]{0.32\textwidth}
        \centering
        \includegraphics[width=0.99\textwidth]{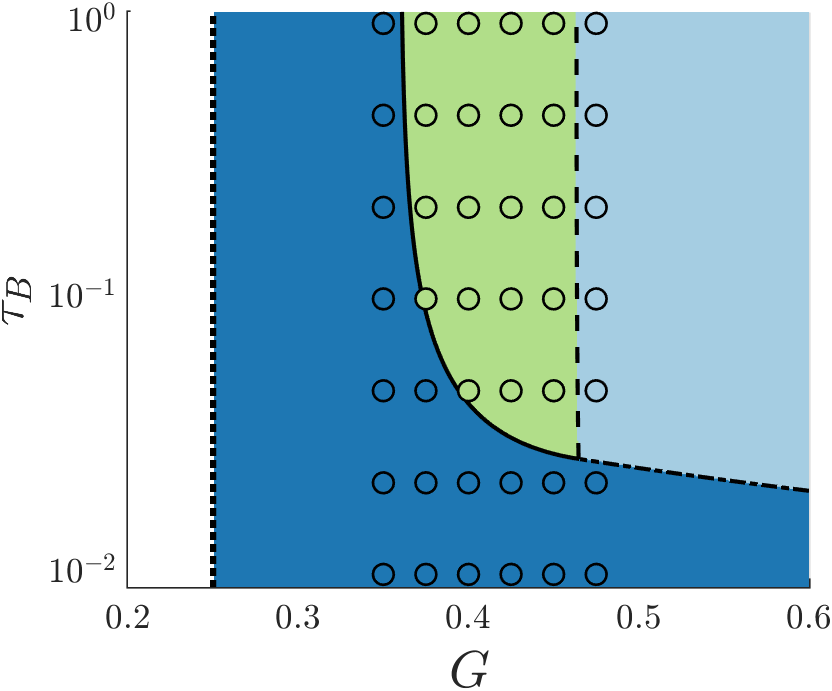}
        \captionsetup{width=0.8\textwidth}
        \caption{$\tau_B \times G$ phase diagram with $\tau_S = 0.001$, $G_B = G_S = 0.4$.}
        \label{fig:phaseDiagram_tauB_G}
    \end{subfigure}
    \begin{subfigure}[t]{0.32\textwidth}
        \centering
        \includegraphics[width=0.99\textwidth]{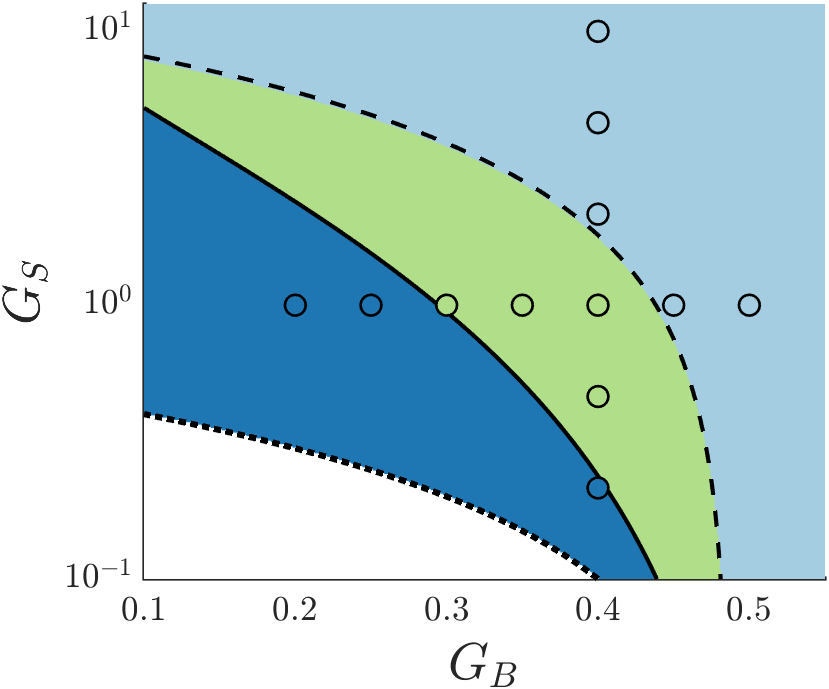}
        \captionsetup{width=0.8\textwidth}
        \caption{$G_B \times G_S$ phase diagram, with $\tau_B = 0.1$, $\tau_S = 0.001$.}
        \label{fig:phaseDiagram_Gb_Gs}
    \end{subfigure}
    \begin{subfigure}[t]{0.32\textwidth}
        \centering
         \includegraphics[width=0.99\textwidth]{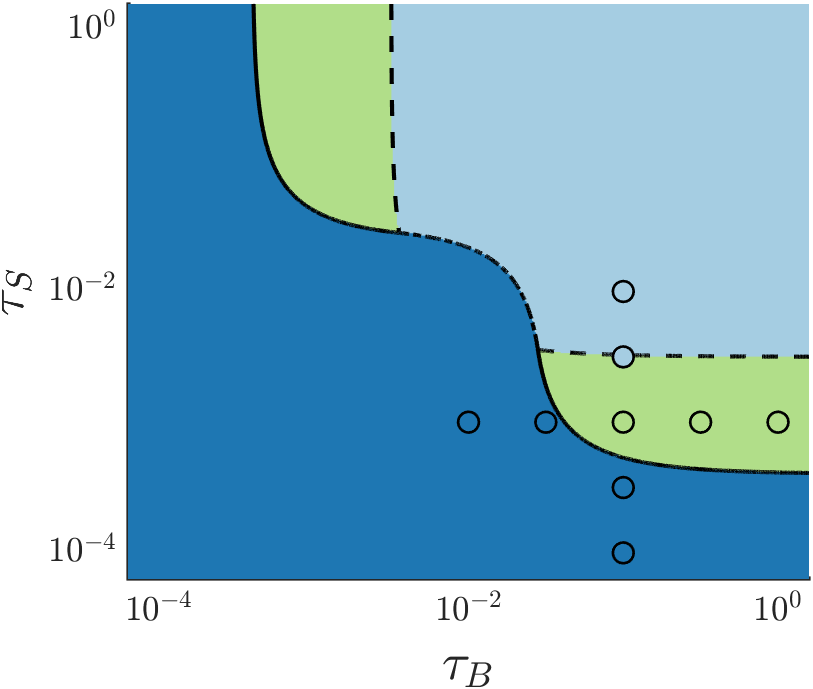}
         \captionsetup{width=0.8\textwidth}
         \caption{$\tau_B \times \tau_S$ phase diagram with $G_B = G_S = 0.4$.}
         \label{fig:phaseDiagram_tauB_tauS}
    \end{subfigure}
    \begin{subfigure}{0.45\textwidth}
        \includegraphics[width = 0.99\textwidth]{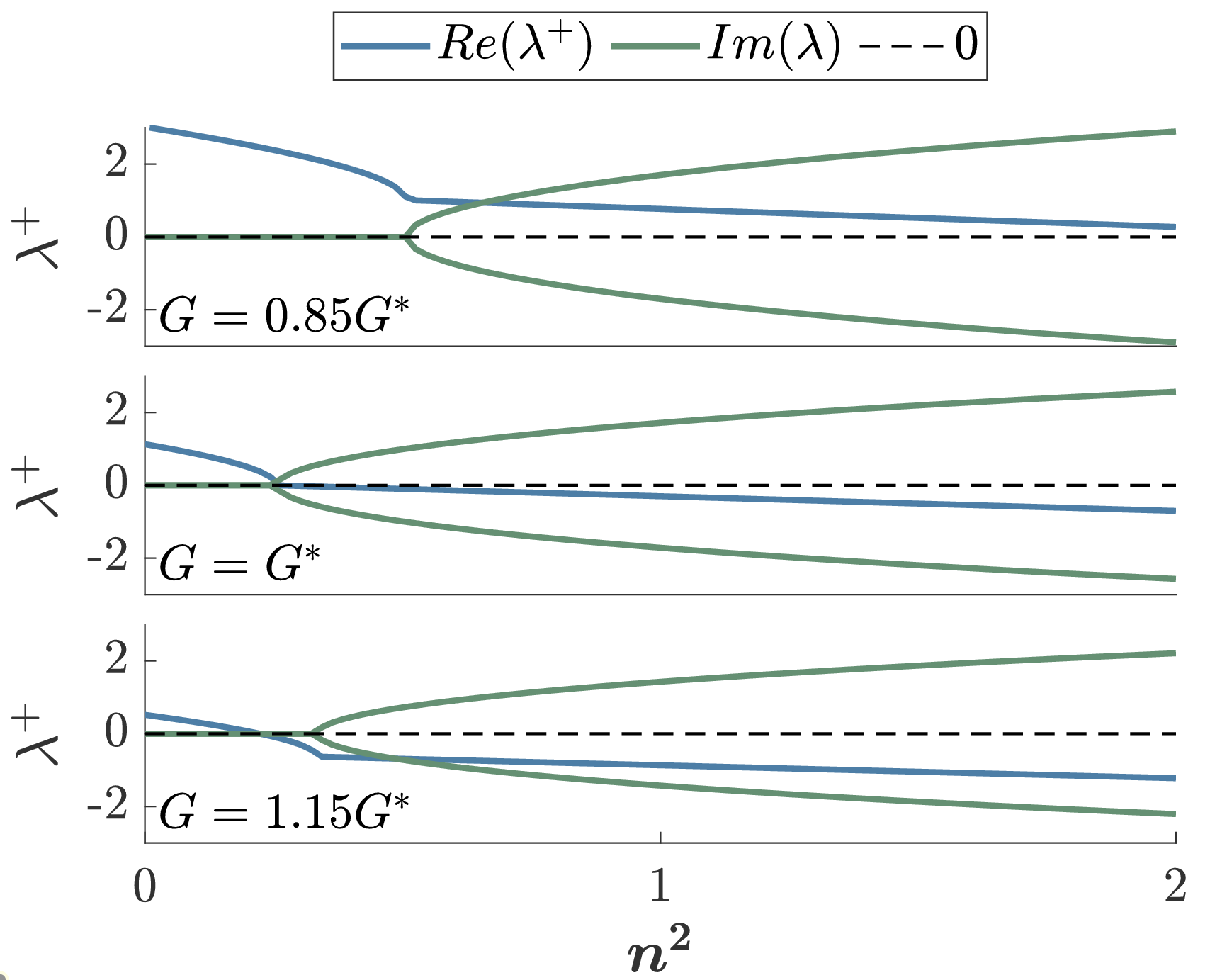}
        \caption{$\text{Re}(\lambda^+)$ and $\text{Im}(\lambda)$ w.r.t. $\vv{n}^2$.}
        \label{fig:eigenValuePicture}
    \end{subfigure}
    \begin{subfigure}{0.45\textwidth}
        \centering
        \includegraphics[width = 0.99\textwidth]{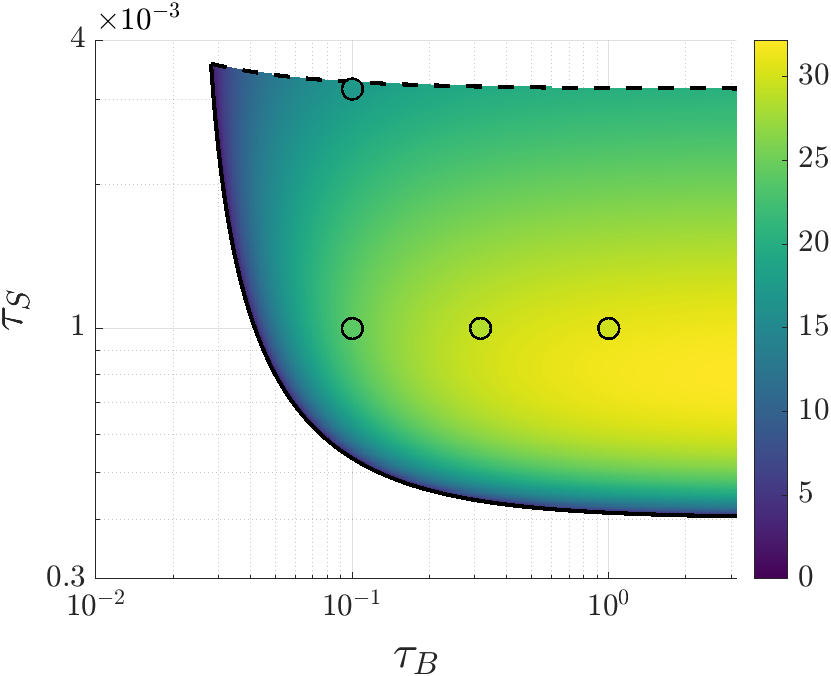}
        \caption{Frequency of the 1-mode oscillations w.r.t. $\tau_B$, $\tau_S$ with $G_S=G_B=0.4$.}
        \label{fig:frequency}
    \end{subfigure}
    \caption{Dynamics in the linear regime. \textbf{(a)-(c)} Three phase diagrams describing the dynamics of the system for different parameter values. The background colour describes the stability according to the linear stability analysis:  white if $G_B + G_S > f'(1)$, dark blue if the $1$-mode is unstable and does not oscillate, light green if the $1$-mode is unstable and oscillates, light blue if all modes are stable. The colour in the circles describes the dynamics according to 2D simulations with these parameter values. They are classified by their values for the correlation coefficient with the 1-modes and the maximal concentration difference on the domain, more details are given in App.~\ref{app:simClassification}. \textbf{(d)} The largest real eigenvalue $\text{Re}(\lambda^+)$ and the imaginary parts $\text{Im}(\lambda^\pm)$ w.r.t. $\vv{n}^2$ according to the linear stability analysis. $\tau_B = 0.1$, $\tau_S = 0.001$, $G_B = G_S = G$. If $G = G^*$ then $\lambda(k^-) = 0$. If $G$ is larger/smaller then $\lambda(k^-)$ is smaller/larger than zero. \textbf{(e)} Frequency of the 1-mode oscillations with respect to the dimensionless relaxation times. The background colour represents the analytical frequency obtained from the linear stability analysis, the colour in the circles the frequency measured in the simulations. The lines in \textbf{(a)-(c)} and \textbf{(e)} are analytically calculated boundaries. The dotted line for Eq.~\eqref{eq:mainCondition}, the dashed line for Eq.~\eqref{eq:b<0}, the dash-dot line for Eq.~\eqref{eq:c>0} and the full line for $\text{Im}\left(\lambda^{\pm}_{|\vv{n}^2 = 1}\right)=0$, which is to indicate where the 1-mode starts to oscillate. The time domain for the simulations is $t \in [0, 0.1]$. The code to generate these figures is found in the SI.}
    \label{fig:phaseDiagrams}
\end{figure}

To study how the parameters affect the behaviour of the system, we generate three phase diagrams and compare the analytic results with 2D numerical simulations (the code to generate them is found in the SI). A time interval of $[0, 0.1]$ is simulated, but the solutions are only considered for the time they are in the linear regime. As initial condition, we choose $c(\vv{x} = \vv{x}_i, t=0) = 1 + r_i$ where $r_i$ is a uniformly distributed random number in $[-10^{-4}, 10^{-4}]$. All other variables are initially zero. A description of the numerical model and how the simulations are classified is found in Apps.~\ref{app:implementation} and \ref{app:simClassification}. The solutions can be divided into three categories; stable, unstable without oscillations and unstable with oscillations. The category for each choice of parameters is expressed by the background colour in \mbox{Figs. \ref{fig:phaseDiagram_tauB_G}-\subref{fig:phaseDiagram_tauB_tauS}}. The results of the simulations are expressed by the colour of the dots in the phase diagrams in Fig.~\ref{fig:phaseDiagrams}. They coincide with the linear stability analysis. The three phase diagrams that are chosen to describe the results of the linear stability analysis are 
\begin{enumerate}
    \item[i)] $\tau_B \times G$ phase diagram, with $G_B = G_S = G$ and $\tau_S = 0.001$ (Fig.~\ref{fig:phaseDiagram_tauB_G}). This gives one shear viscous component and a bulk component ranging from viscous to elastic. As expected, increasing the elastic modulus $G$ makes the system more stable. A minimal bulk relaxation time $\tau_B$ is also needed for oscillations. 
    \item[ii)] $G_B \times G_S$ phase diagram, with $\tau_B = 0.1$ and $\tau_S = 0.001$ (Fig.~\ref{fig:phaseDiagram_Gb_Gs}), so that there is both a viscous and an elastic component. The bulk elastic modulus has more influence on the dynamics. This implies that the dynamics are mainly decided by the elastic component. 
    \item[iii)] $\tau_B \times \tau_S$ phase diagram, with $G_B = G_S = 0.4$ (Fig.~\ref{fig:phaseDiagram_tauB_tauS}). We observe that the phase diagram is mirrored in the $\tau_B = \tau_S$ line, demonstrating how $\tau_B$ and $\tau_S$ can be interchanged in the case of equal elastic moduli. Additionally, there is a distance between the $\tau_B =  \tau_S$ line and the parameters that result in oscillations. This indicates that significantly different relaxation times are needed for oscillations. 
\end{enumerate}
To demonstrate which modes are dominant, we show the eigenvalue w.r.t. $\vv{n}^2$ for three different sets of parameters in Fig.~\ref{fig:eigenValuePicture}. The value of $G^*$ is chosen such that $\lambda_{|\vv{k}^2 = k^-} = 0$ if $G = G^*$, which is defined as $G^* = \frac{f'(1)(\tau_B^2 + \tau_S^2)}{(\tau_B + \tau_S)^2}$. It is important to note that the largest real part $\text{Re}(\lambda^+)$ is decreasing, as is shown in App.~\ref{app:linStabAna}.

We calculate the frequency of the oscillations as the imaginary part of the eigenvalue (the code to do so is found in the SI). In Fig.~\ref{fig:frequency}, this frequency is shown in dependence of the relaxation times. Here $G_B=G_S = 0.4$ and $\tau_B > \tau_S$, which gives a more solid-like bulk and a more liquid-like shear stress dynamics. It is noteworthy that when increasing $\tau_B$, the frequency increases and converges to a constant value, see Fig.~\ref{fig:frequency}. From this, we conclude that the oscillations are not caused by the relaxation of the elastic stress, because in that case the frequency should reduce to zero when $\tau_B$ is increased.

\begin{figure}
    \centering
    \begin{subfigure}{\textwidth}
        \includegraphics[width=0.16\textwidth]{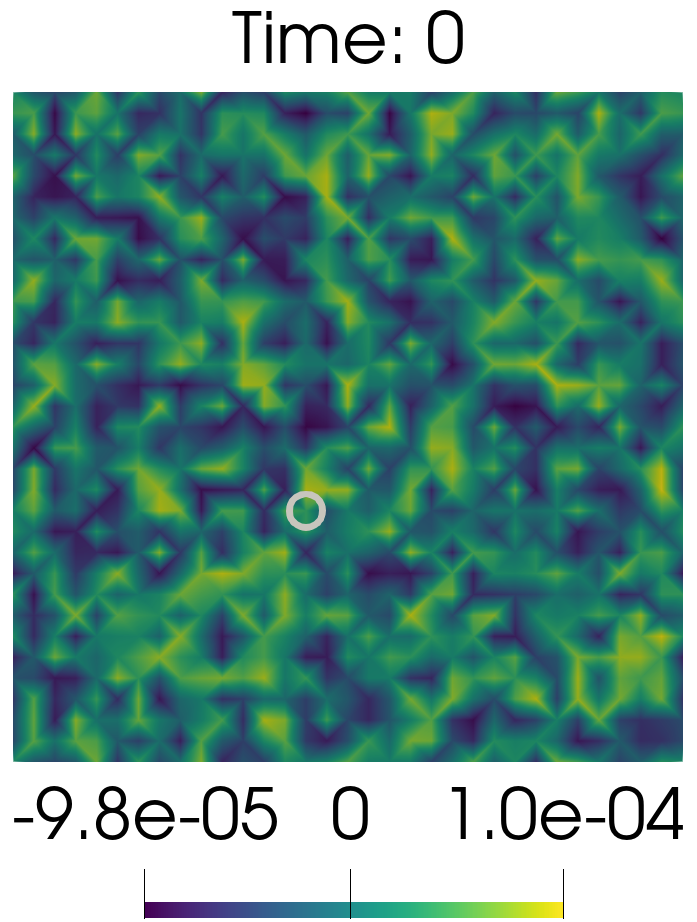}
        \includegraphics[width=0.16\textwidth]{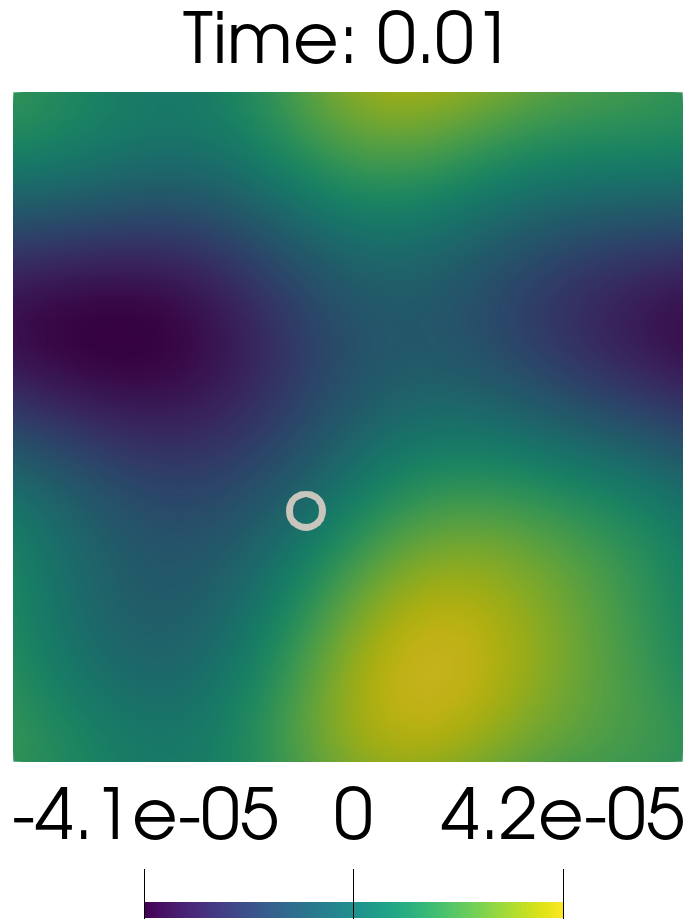}
        \includegraphics[width=0.16\textwidth]{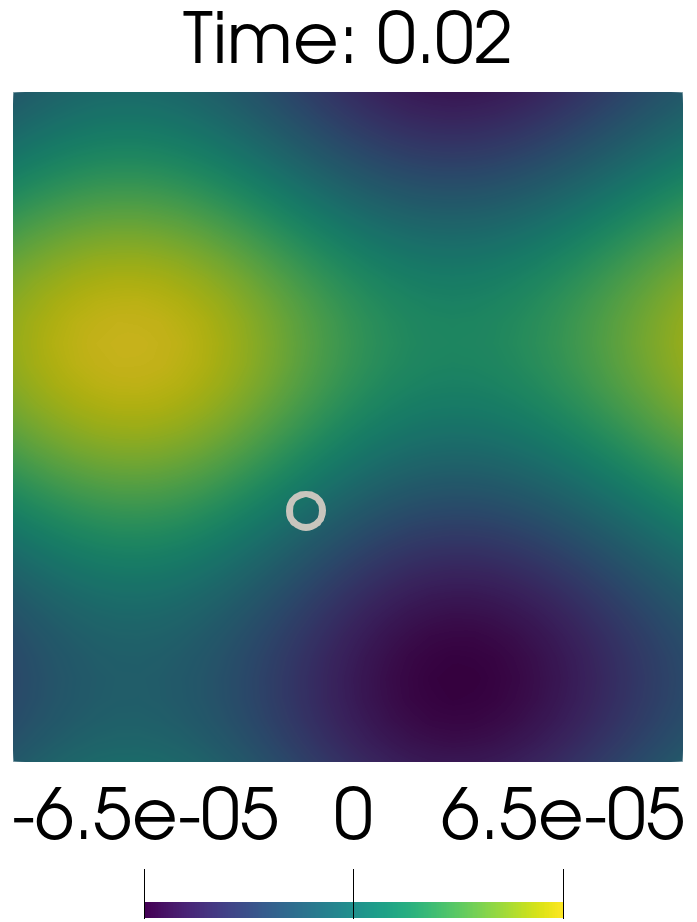}
        \includegraphics[width=0.16\textwidth]{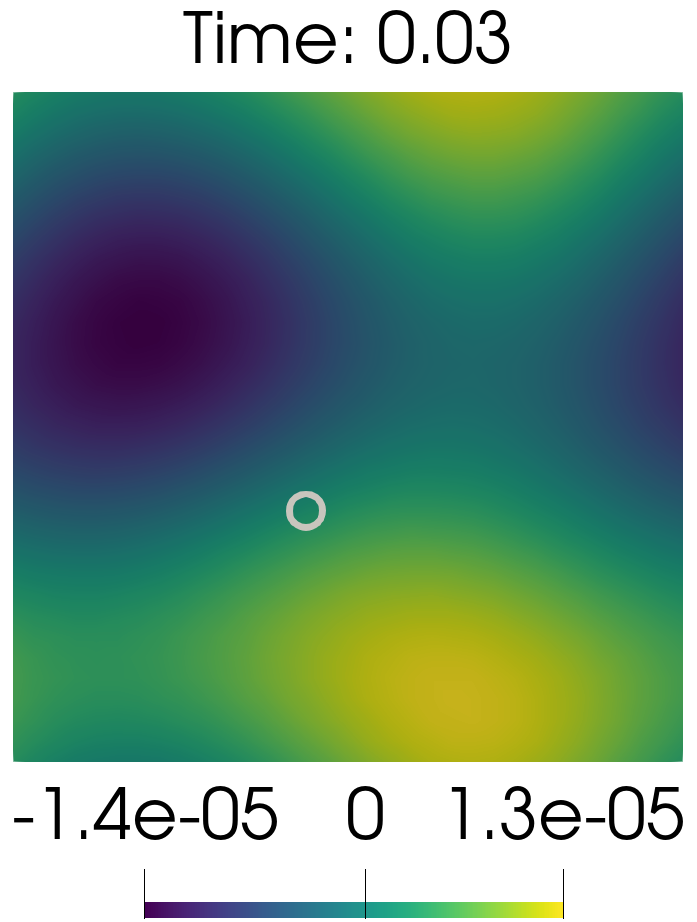}
        \includegraphics[width=0.16\textwidth]{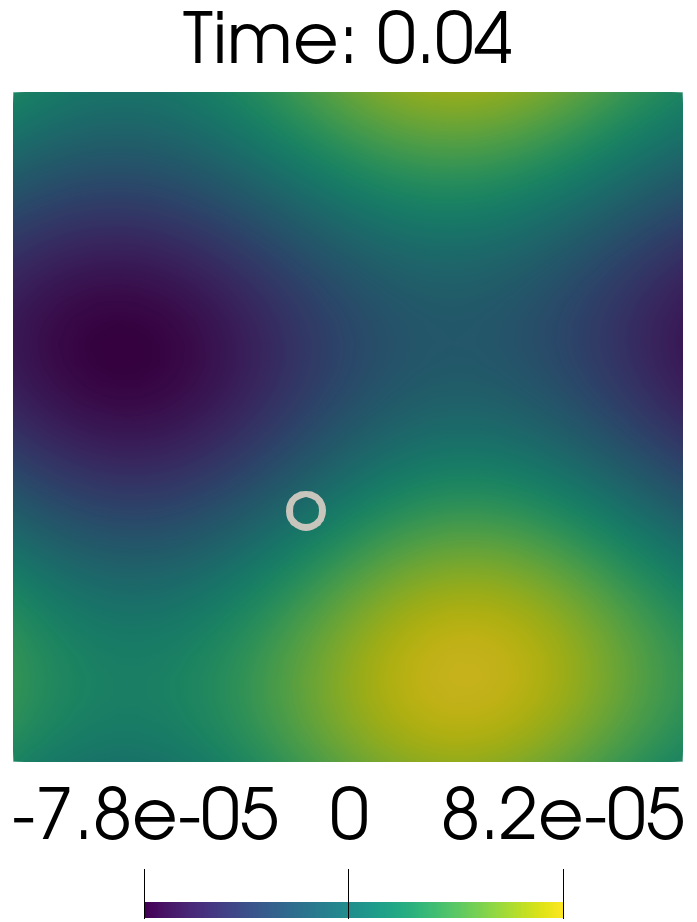}
        \includegraphics[width=0.16\textwidth]{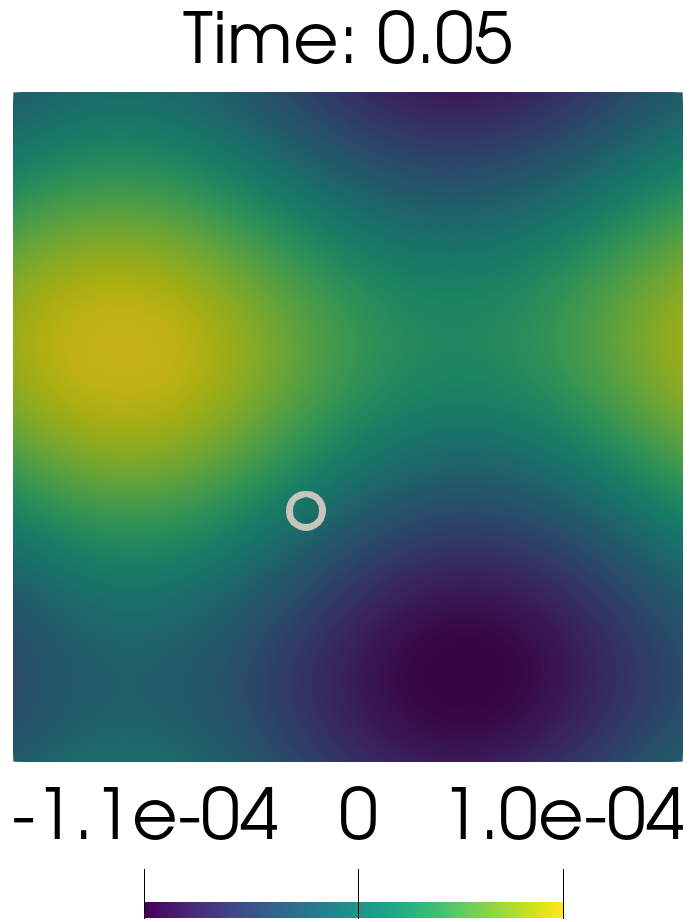}
        \caption{$c-1$ for times $0, 0.01, ..., 0.05$}
        \label{fig:2D_c_plot}
    \end{subfigure}
    \begin{subfigure}[b]{0.49\textwidth}
        \includegraphics[width=0.99\textwidth]{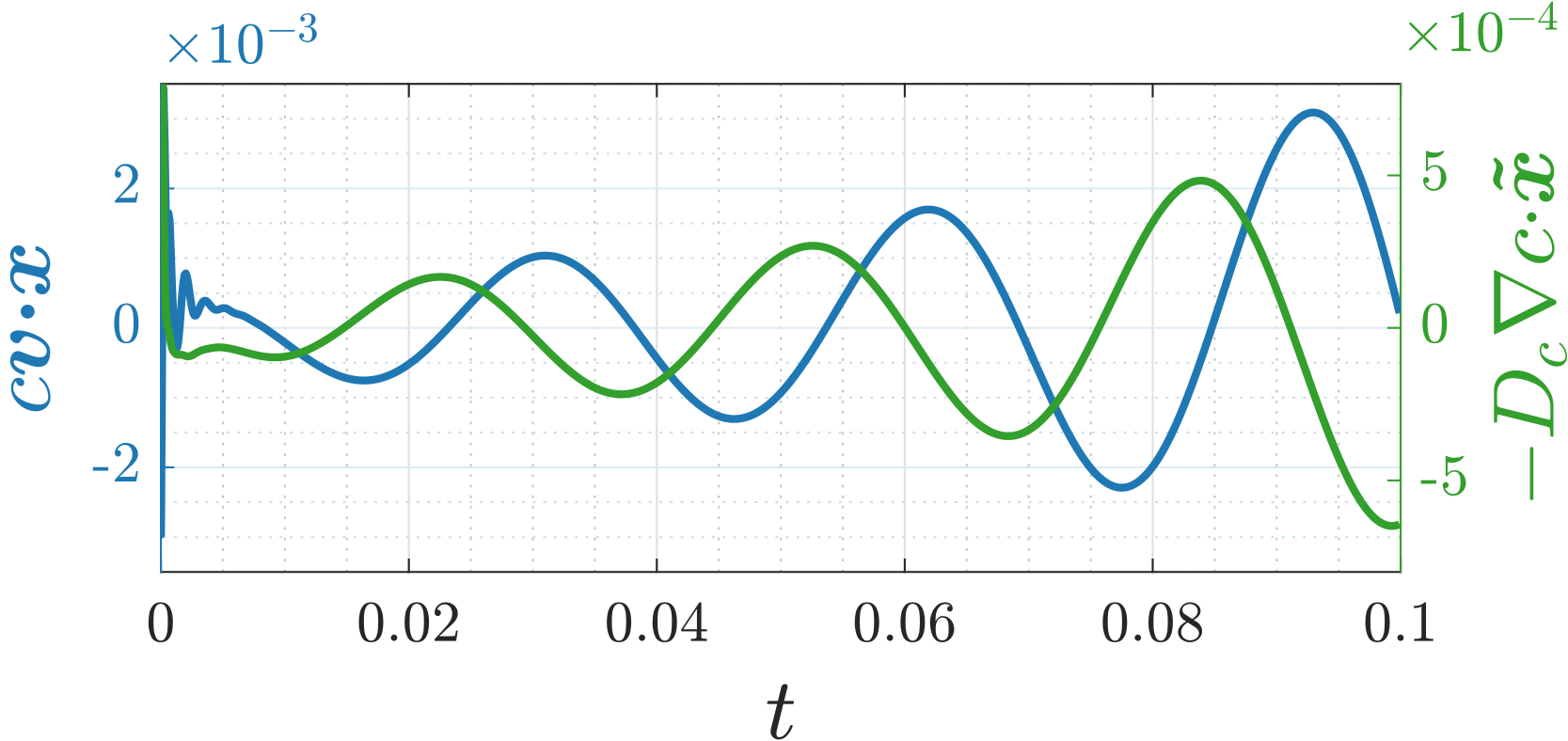}
        \caption{Advective and diffusive flux}
        \label{fig:fluxes}
    \end{subfigure}
    \begin{subfigure}[b]{0.49\textwidth}
        \includegraphics[width=0.99\textwidth]{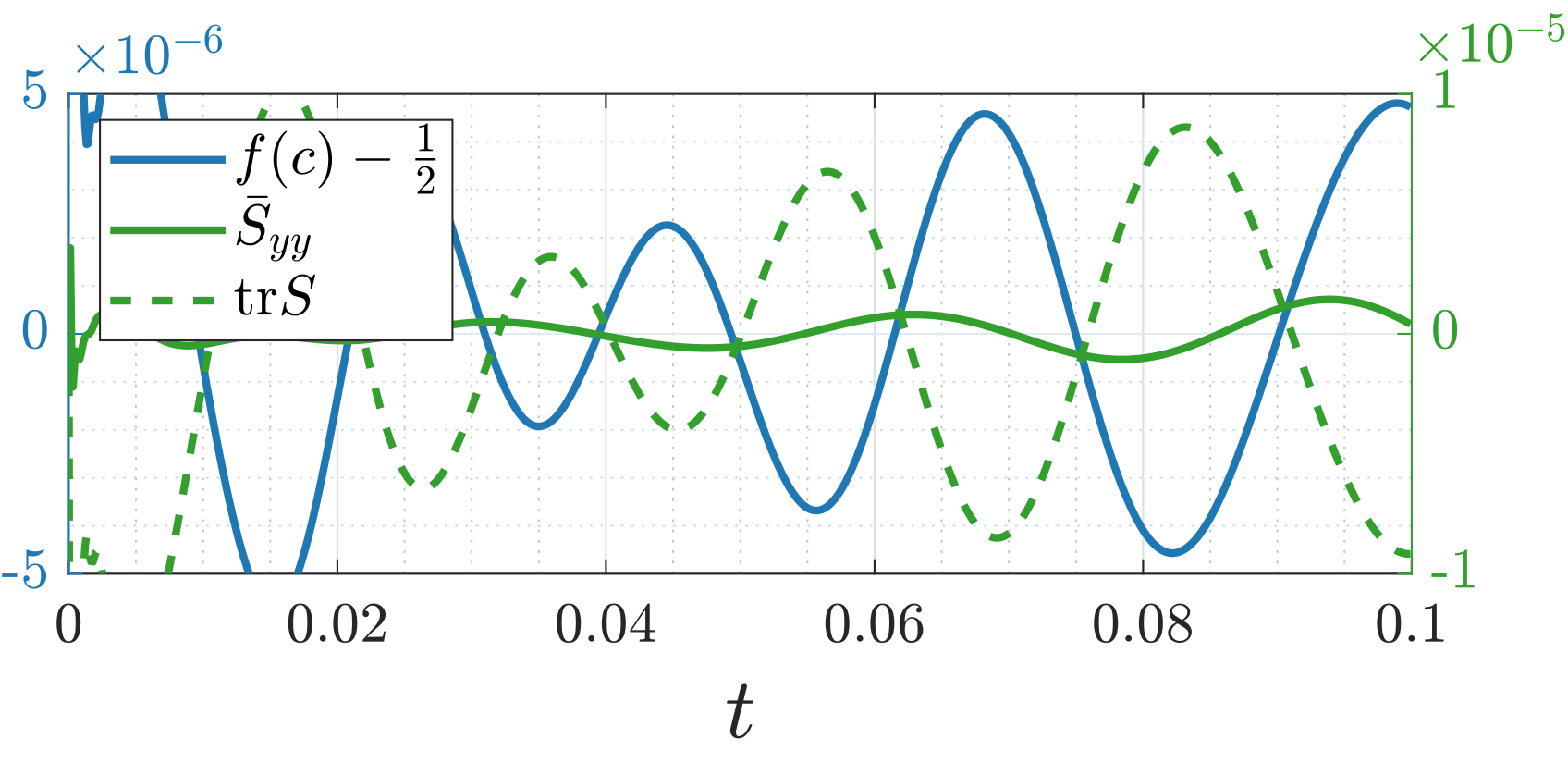}
        \caption{Active, elastic and viscous stress contributions}
        \label{fig:allStressOverTime}
    \end{subfigure}
    \caption{Linear dynamics in 2D. \textbf{(a)} Simulation results of the time evolution of the concentration perturbation $c-1$. Snapshots are shown for times $0, 0.01, ..., 0.05$. The dot at $\vv{x} = \vv{x}^* = (-0.0625, -0.125)^T$ indicates the point for which the values in \textbf{(b)} and \textbf{(c)} are plotted. It is chosen to be the point between the maximum and minimum of $c$. \textbf{(b)} Advective and diffusive fluxes  over time, along the line between the minimum and the maximum, i.e. $c\vv{v}\cdot\tilde{\vv{x}}$ and $D_c \nabla c \cdot \tilde{\vv{x}}$ with $\tilde{\vv{x}} = (1/\sqrt{2},-1/\sqrt{2})^T$. \textbf{(c)} Active, elastic and viscous stress contributions, $f(c)$, $\tr (S)$ and $\bar{S}_{yy}$. The used parameters are $G_B = G_S = 0.45$, $\tau_B = 1$, $\tau_S = 0.001$.}
    \label{fig:changeOverTime}
\end{figure}

To better understand the evolution of the system over time, we compare the time dynamics of the fluxes and stresses of a numerical solution. A set of parameters is chosen that corresponds to a dominantly elastic bulk stress, a dominantly viscous shear stress and which yields a complex eigenvalue with a real part  close to zero. The results are shown in Fig.~\ref{fig:changeOverTime}. 
In Fig.~\ref{fig:2D_c_plot}, the field of concentration perturbations $c-1$ is displayed. From the random initial condition, a linear combination of the two 1-modes develops. Then, the concentration starts to oscillate, causing the peak and the valley to swap places over time. To study the underlying dynamics, the convective flux and diffusive flux along the line between the minimum and maximum of $c$ are shown in Fig.~\ref{fig:fluxes}. Both fluxes are shown at $\vv{x} = \vv{x}^* = (-0.0625, -0.125)^T$, which is the point exactly between the two extrema of $c$. So for the diffusive flux, we show $-D_c\nabla c \cdot \tilde{\vv{x}}$ and for the convective flux $c \vv{v} \cdot \tilde{\vv{x}}$, with $\tilde{\vv{x}} = (1/\sqrt{2},-1/\sqrt{2})^T$. The convective and diffusive fluxes are almost perfectly out of phase. When the convective flux is maximal, then the diffusive flux decreases the most. A peak in the advective flux is also followed by a valley in the diffusive flux, so transport of $c$ by advection is followed by diffusion of $c$ in the opposite direction. 
In Fig.~\ref{fig:allStressOverTime}, both the viscoelastic stress and the active stress amplitude $f(c)$ are shown. The viscoelastic stress is represented by the bulk stress $\tr (S)$ and one component of the shear stress tensor $\bar{S}_{yy}$, which is chosen because the non-diagonal entries of $\bar{S}$ are negligible. 

The elastic stress is approximately in phase with the diffusive flux while the active stress is $\approx 180^\circ$ phase-shifted. Therefore, at the time point of maximal elastic bulk deformation (dilation), concentration gradients are maximal and corresponding local concentrations and active stresses are minimal. This also implies that a peak in advective flux is followed by a peak in active stress, indicating that the concentration differences are caused by the advective flux. Not surprisingly, we find that the viscous stress is in phase with the advective flux. 

As a result of the linear stability analysis and the simulations, we found the following explanation for the oscillations. The oscillations require two different relaxation times. So either the bulk or shear component will be more elastic and the other more viscous. A small increase in concentration results in active stress driven advective fluxes towards this area. These fluxes deform the surface, increasing the \mbox{(quasi-)}elastic stresses. However, they also transport the surface bound species, increasing the local concentration and active stress even more. This positive feedback loop continues until the active stress is balanced by the elastic stress. In this moment, the concentration peak is maximal and the advective flux stops. Since the concentration is not homogeneous, diffusive fluxes reduce the concentration peak. Due to the (quasi-)viscous component of the surface, however, the deformation reduces slower and the \mbox{(quasi-)}elastic stress will remain stronger than the active stress. This eventually causes a peak in concentration in a different area and the cycle repeats itself.

\section{Dynamics in the nonlinear regime}
\label{sec:nonlin}
\begin{figure}
    \includegraphics[width=0.16\textwidth]{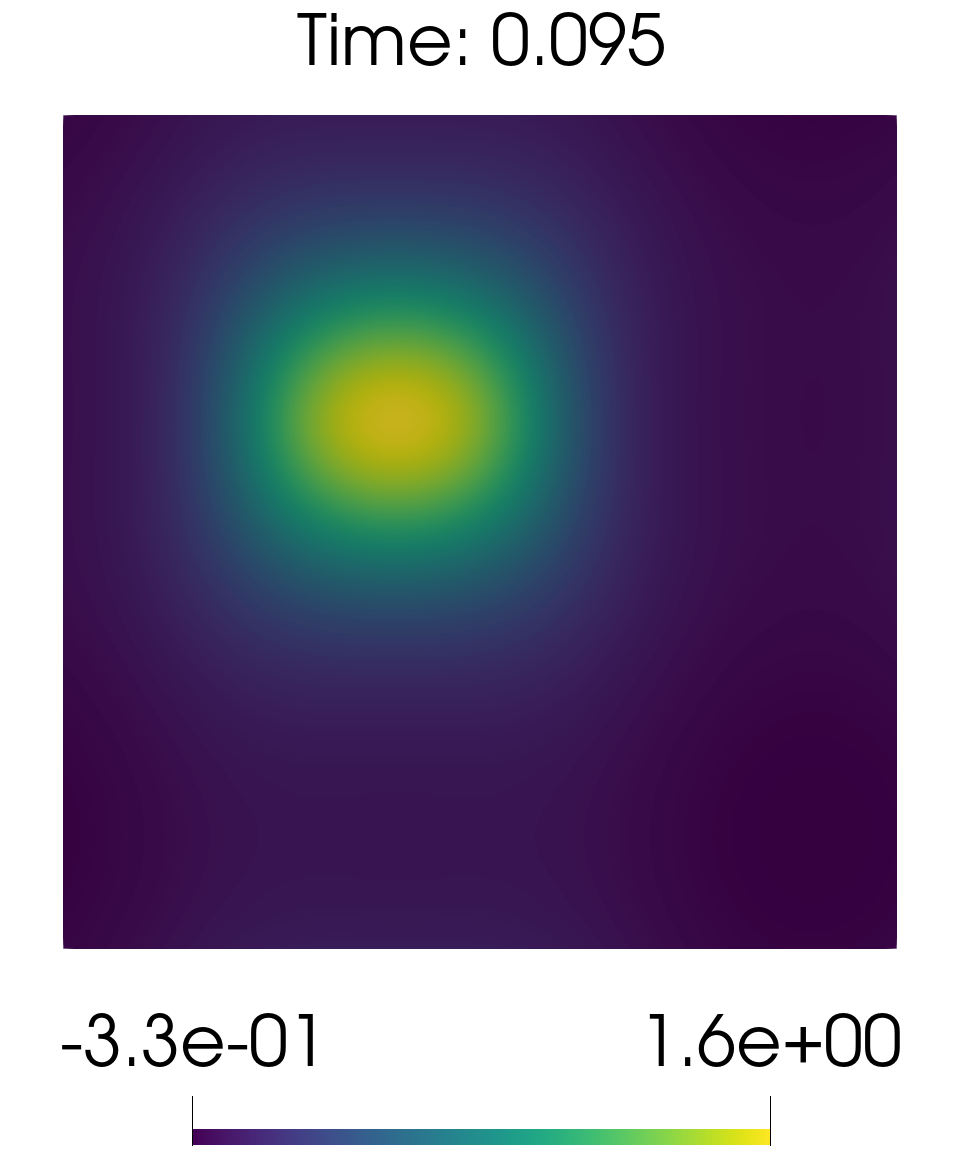}
    \includegraphics[width=0.16\textwidth]{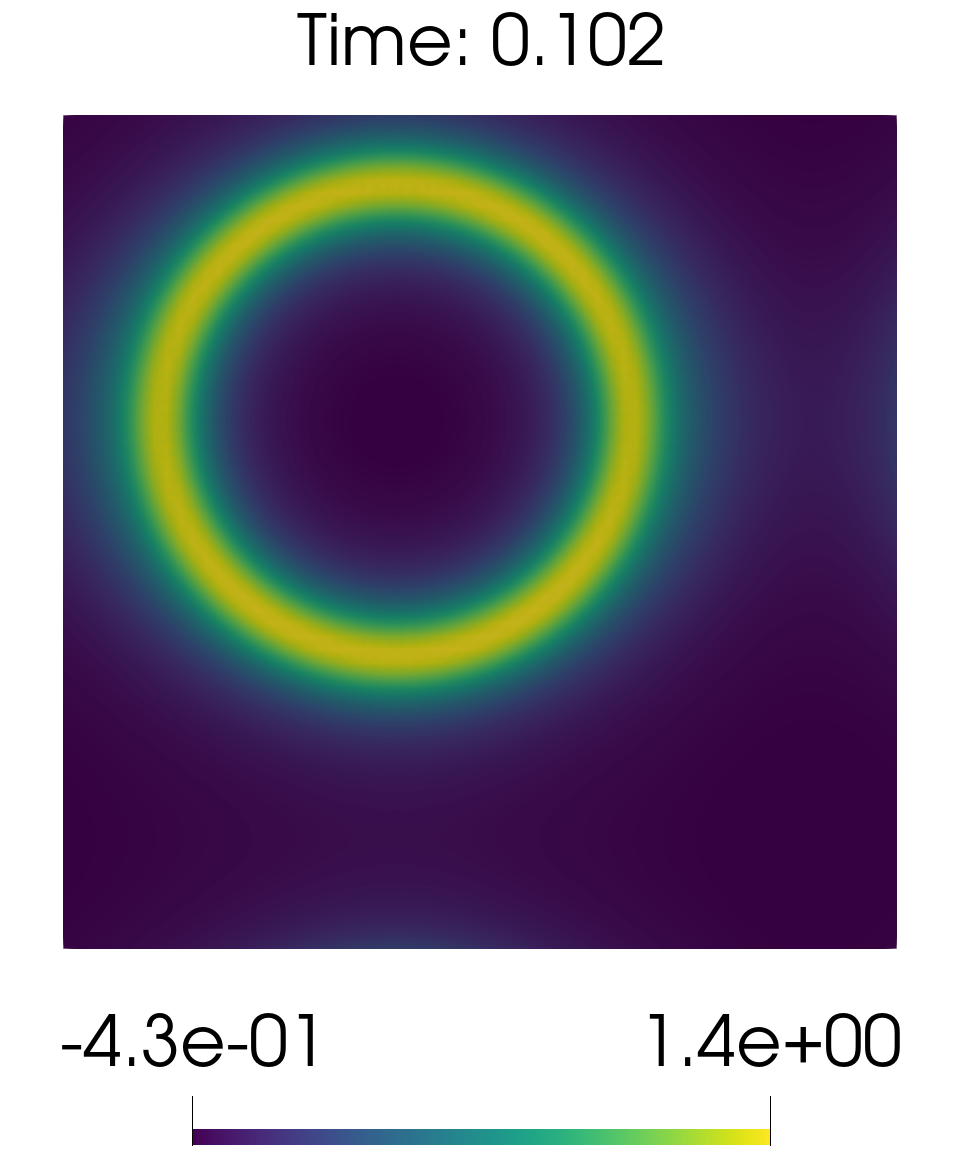}
    \includegraphics[width=0.16\textwidth]{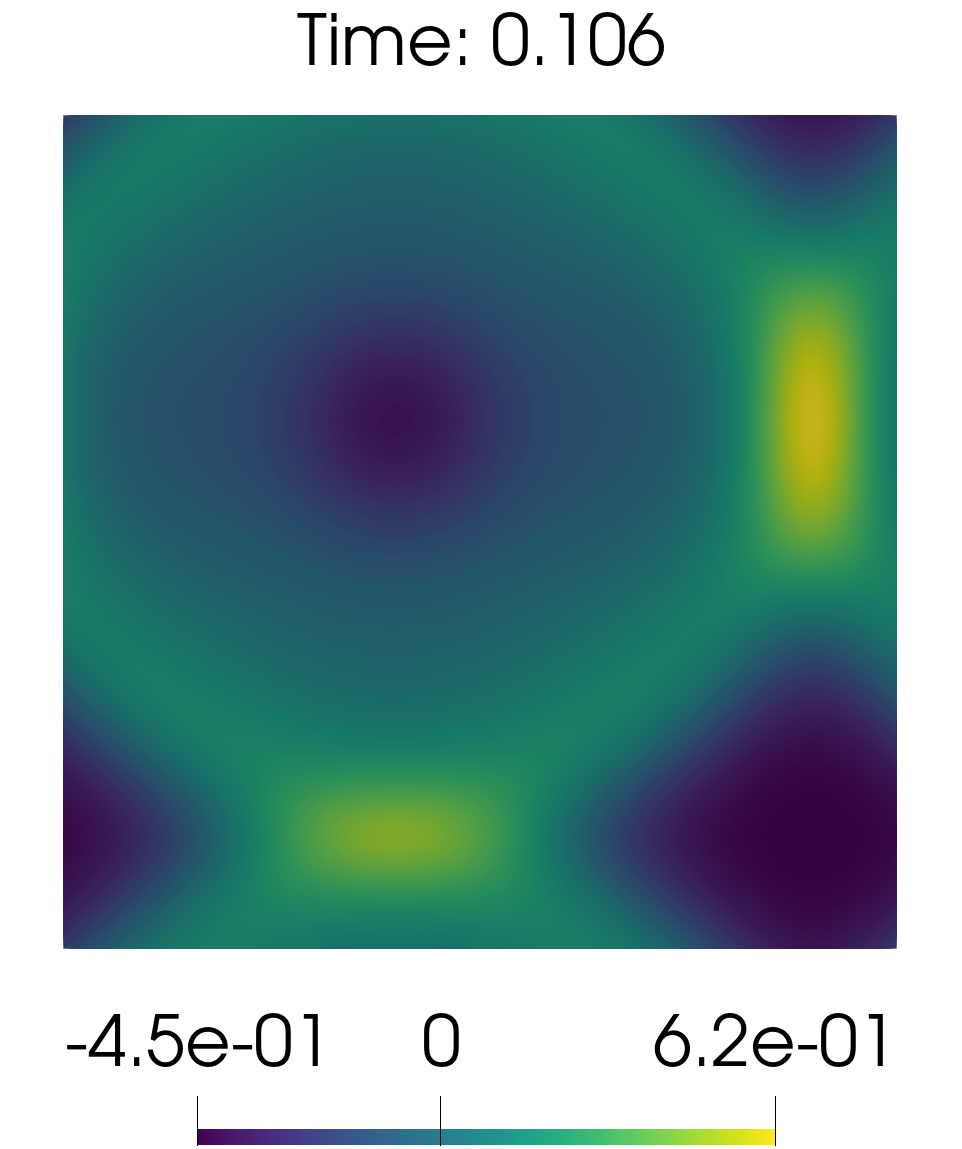}
    \includegraphics[width=0.16\textwidth]{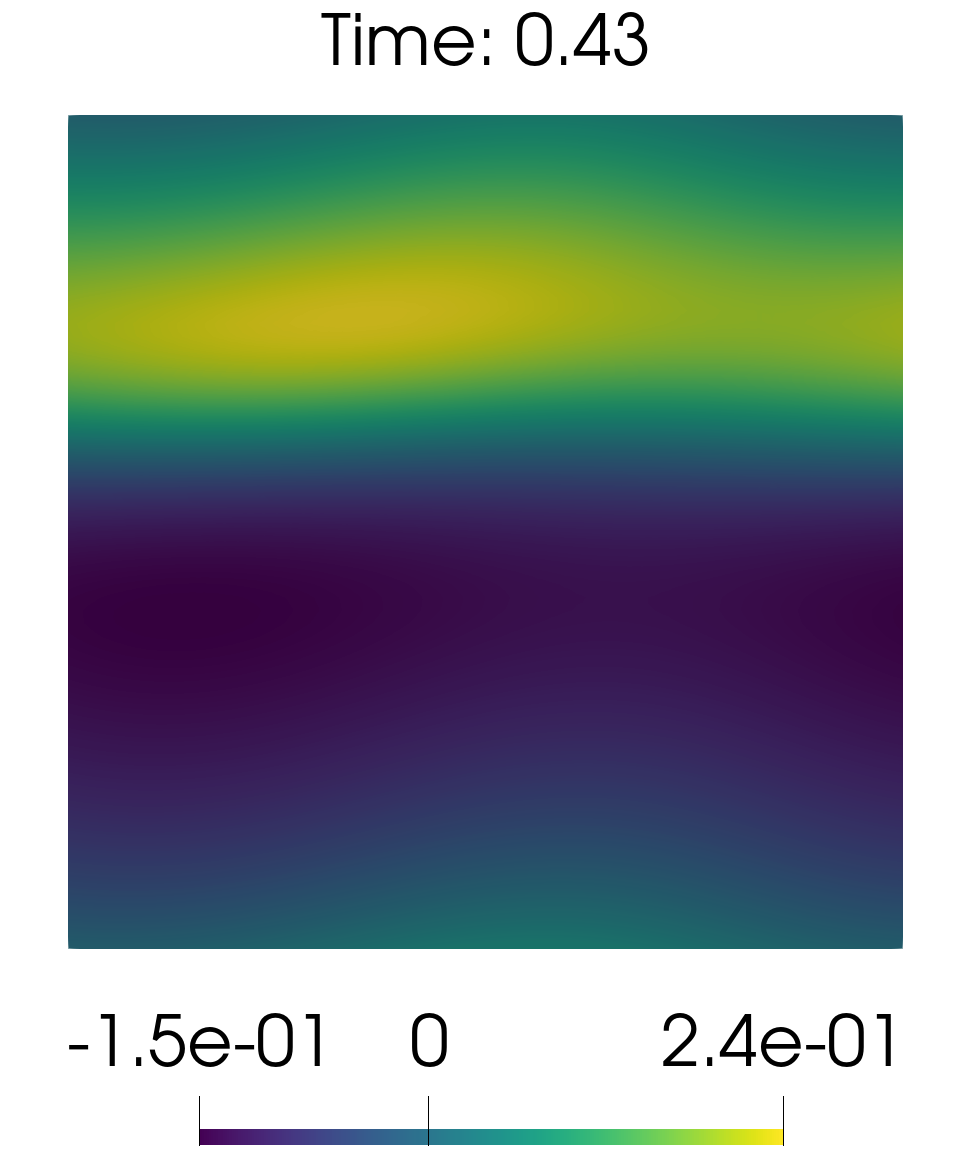}
    \includegraphics[width=0.16\textwidth]{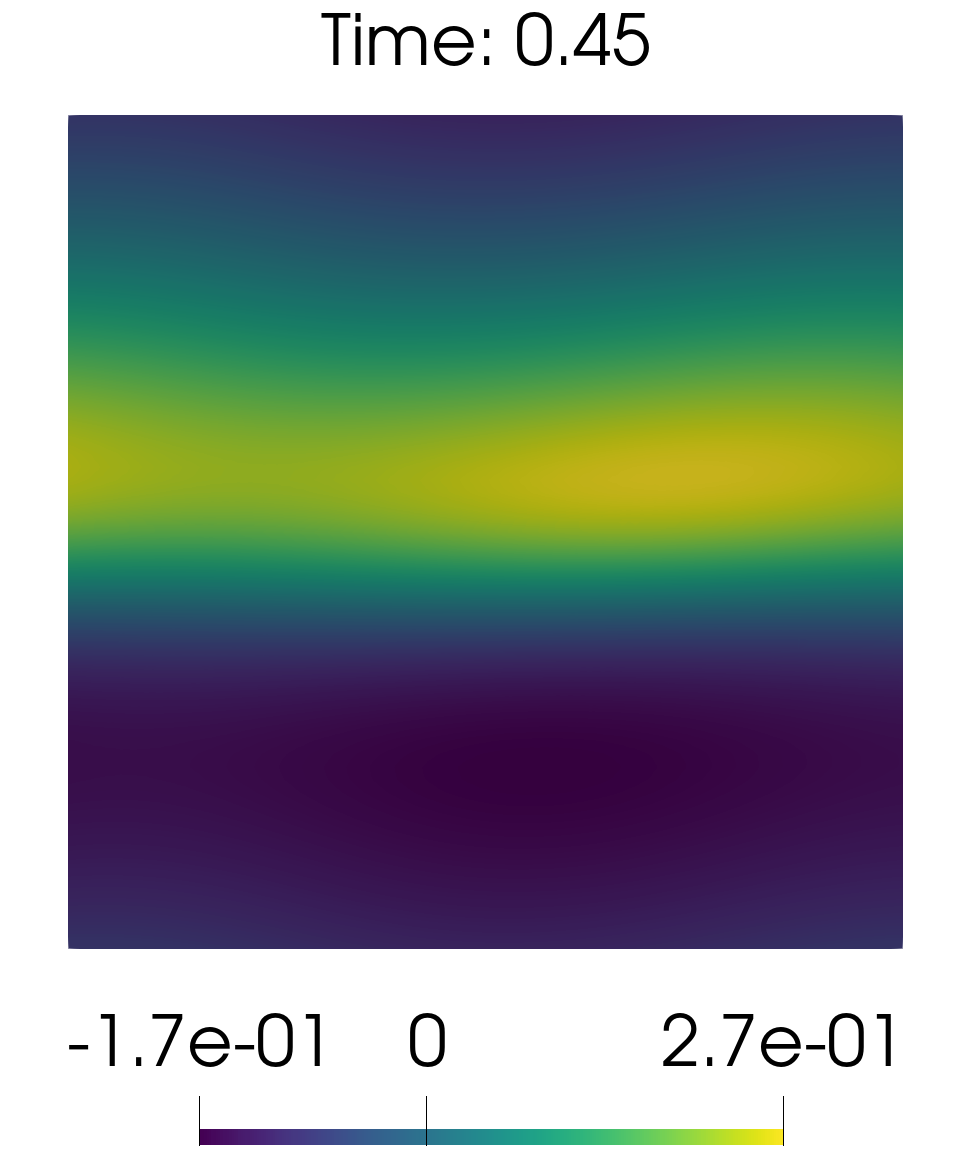}
    \includegraphics[width=0.16\textwidth]{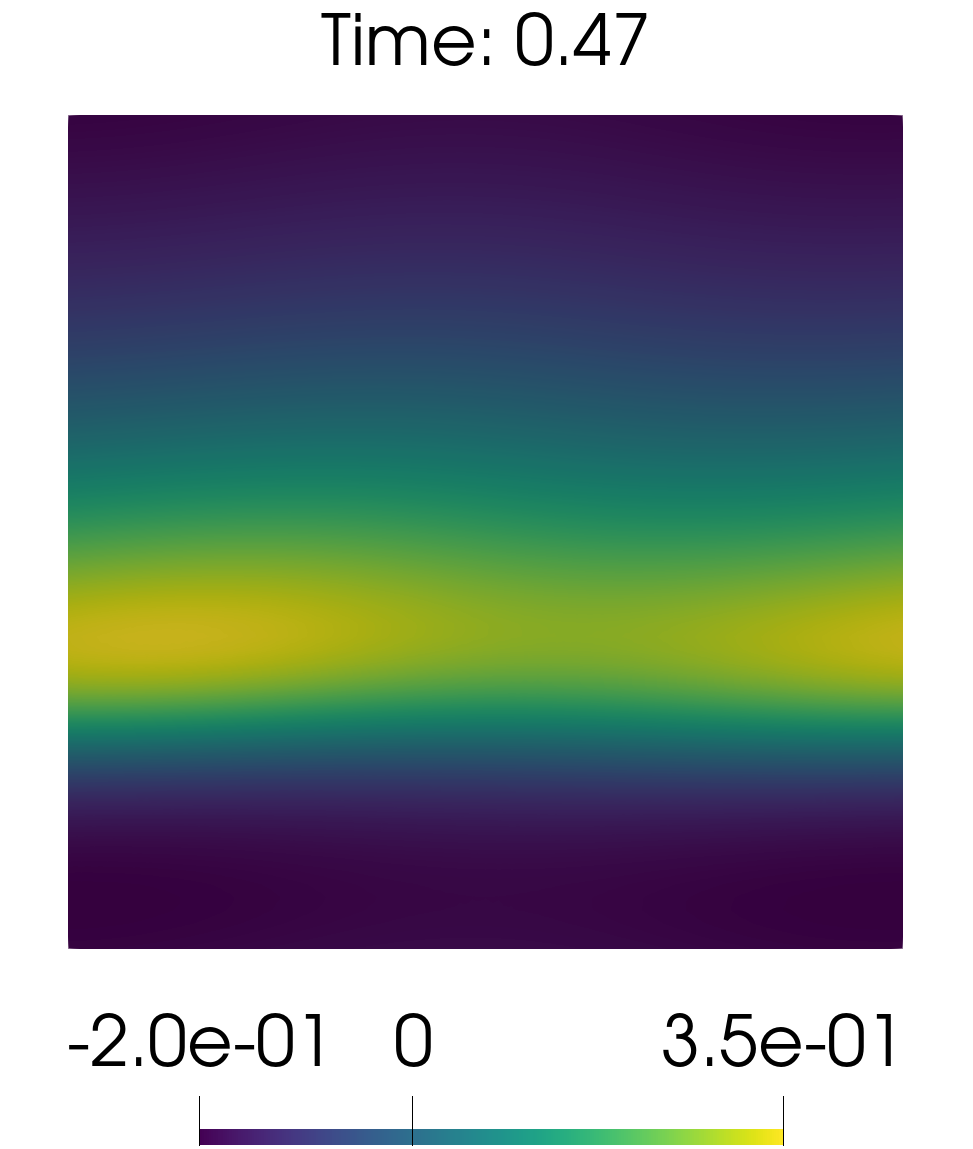}
    \caption{{Nonlinear dynamics in 2D.} Panels show the concentration perturbation $c-1$ for several points in time corresponding to movie 1 in SI. The parameters are $\tau_B = 0.1$, $\tau_S = 0.001$, $G_B = G_S = 0.4$. The first three time points show the transition from standing to travelling waves. The last three show a travelling wave with oscillating amplitude.}
    \label{fig:nonLin2D}
\end{figure}

Using numerical simulations, the system can also be studied far from equilibrium. In particular, cases with $G_B + G_S < f'(1)$ can also be simulated. The numerical solutions show that the nonlinear terms in the system of equations (Eqs.~\eqref{eq:devS_scaled}-\eqref{eq:force_scaled}) have a stabilising effect on the dynamics such that the linearly unstable cases do not diverge. In Fig.~\ref{fig:nonLin2D} (movie 1 in SI), the concentration $c-1$ is shown for $\tau_B = 0.1$, $\tau_S = 0.001$, $G_B = G_S = 0.4$. At the start of the nonlinear regime the pattern remains as predicted by the linear stability analysis. Then the peak in $c$ continues to increase, until it changes into an outward travelling wave with the location of the original peak as epicentre, $t\leq 1.06$ in Fig.~\ref{fig:nonLin2D}. Initially, complicated dynamics develop from travelling waves, crossing the boundary and colliding with each other. The maximal difference in concentration (defined in the appendix in Eq.~\eqref{eq:deltaCmax}) oscillates with a frequency similar to the frequency predicted by the linear stability analysis. But, these oscillations slowly diminish and are replaced by one with a different frequency. Eventually a consistent pattern develops of a travelling wave with an oscillating amplitude, see Fig.~\ref{fig:nonLin2D}, $t \geq 0.2$. 

\begin{figure}
    \begin{subfigure}{\textwidth}
        \includegraphics[width = 0.49\textwidth, valign = t]{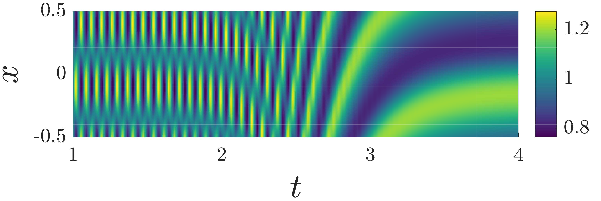}
        \includegraphics[width = 0.49\textwidth, valign = t]{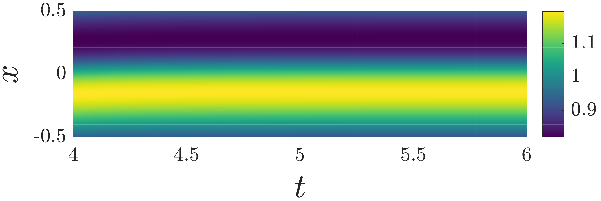}
        \vspace{-0.5cm}
        \caption{$G_B = G_S= 0.45$, $\tau_B = 10^{-1.5}$, $\tau_S = 0.001$.}
        \label{fig:kymoStat}
    \end{subfigure}
    \centering
    \begin{subfigure}{\textwidth}
        \includegraphics[width = 0.49\textwidth, valign = t]{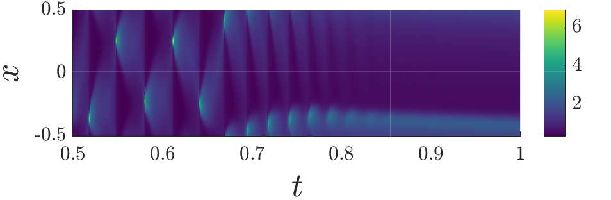}
        \includegraphics[width = 0.49\textwidth, valign = t]{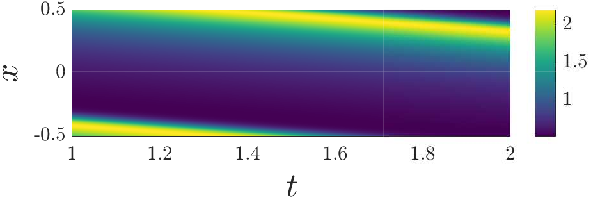}
        \vspace{-0.5cm}
        \caption{$G_B = G_S = 0.2$, $\tau_B = 10^{-1.5}$, $\tau_S = 0.001$.}
        \label{fig:kymoG02}
    \end{subfigure}
    \begin{subfigure}{\textwidth}
        \includegraphics[width = 0.49\textwidth, valign = t]{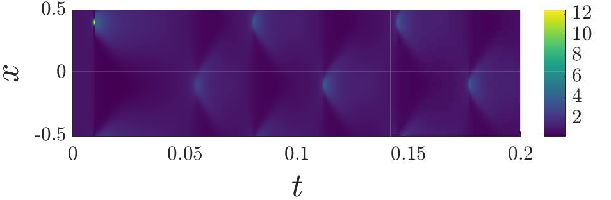}
        \includegraphics[width = 0.49\textwidth, valign = t]{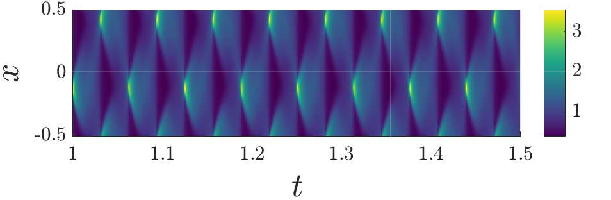}
        \vspace{-0.5cm}
        \caption{$G_B = G_S = 0.3$, $\tau_B = 10^{-1.5}$, $\tau_S = 0.001$.}
        \label{fig:kymoG03}
    \end{subfigure}
    \begin{subfigure}{\textwidth}
        \includegraphics[width = 0.49\textwidth, valign = t]{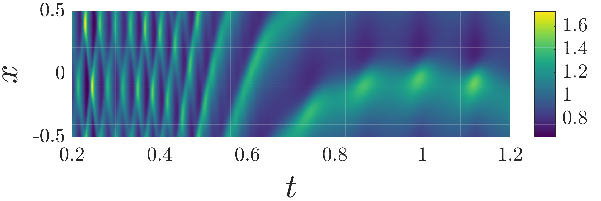}
        \includegraphics[width = 0.49\textwidth, valign = t]{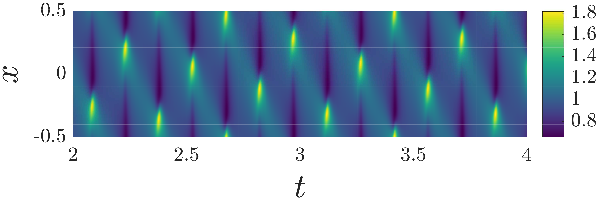}
        \vspace{-0.5cm}
        \caption{$G_B = G_S = 0.4$, $\tau_B = 0.1$, $\tau_S = 0.001$.}
        \label{fig:nonLinKymo}
    \end{subfigure}
    \caption{Kymographs of the concentration field $c$ illustrate oscillatory patterns and pattern transitions in the nonlinear regime. Left: development of the consistent nonlinear dynamics. Right: the consistent nonlinear dynamics at late times.}
    \label{fig:kymographs}
\end{figure}

\begin{figure}
    \begin{subfigure}[t]{0.32\textwidth}
        \captionsetup{width=0.9\textwidth}
        \includegraphics[width = 0.99\textwidth]{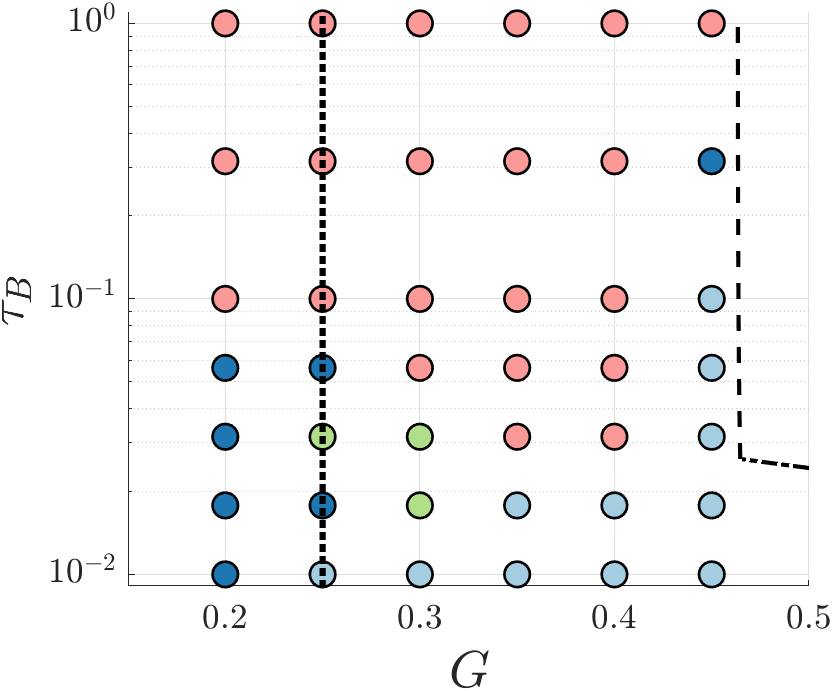}
        \caption{$\tau_B \times G$ phase diagram, $G_B = G_S = G$, $\tau_S = 0.001$}
        \label{fig:nonLinClassTauB_G}
    \end{subfigure}
    \begin{subfigure}[t]{0.33\textwidth}
        \includegraphics[width = 0.99\textwidth]{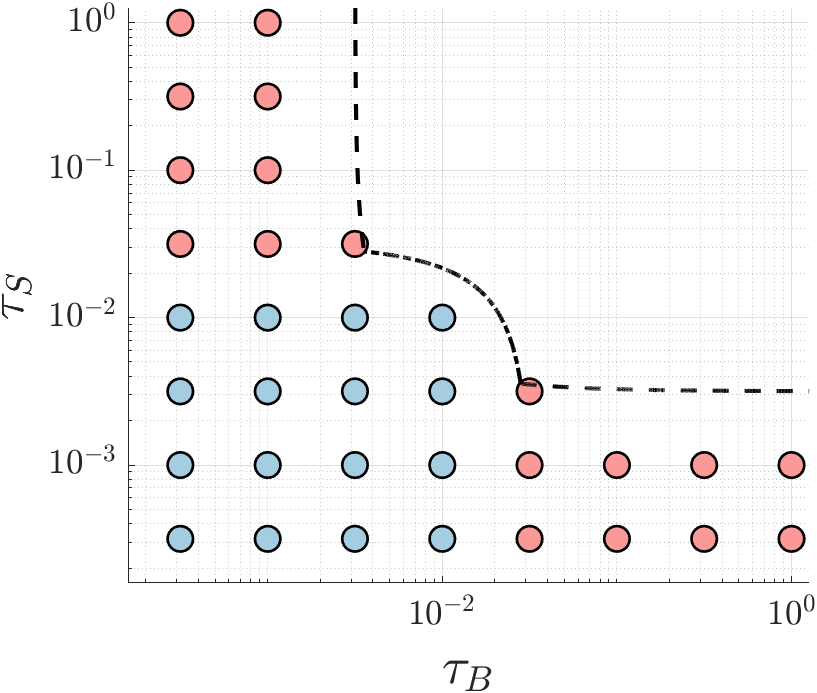}
        \caption{$\tau_B \times \tau_S$ phase diagram, $G_B = G_S = 0.4$}
        \label{fig:nonLinClassTauB_TauS}
    \end{subfigure}
    \begin{subfigure}[t]{0.32\textwidth}
        \captionsetup{width=0.8\textwidth}
        \includegraphics[width = 0.99\textwidth]{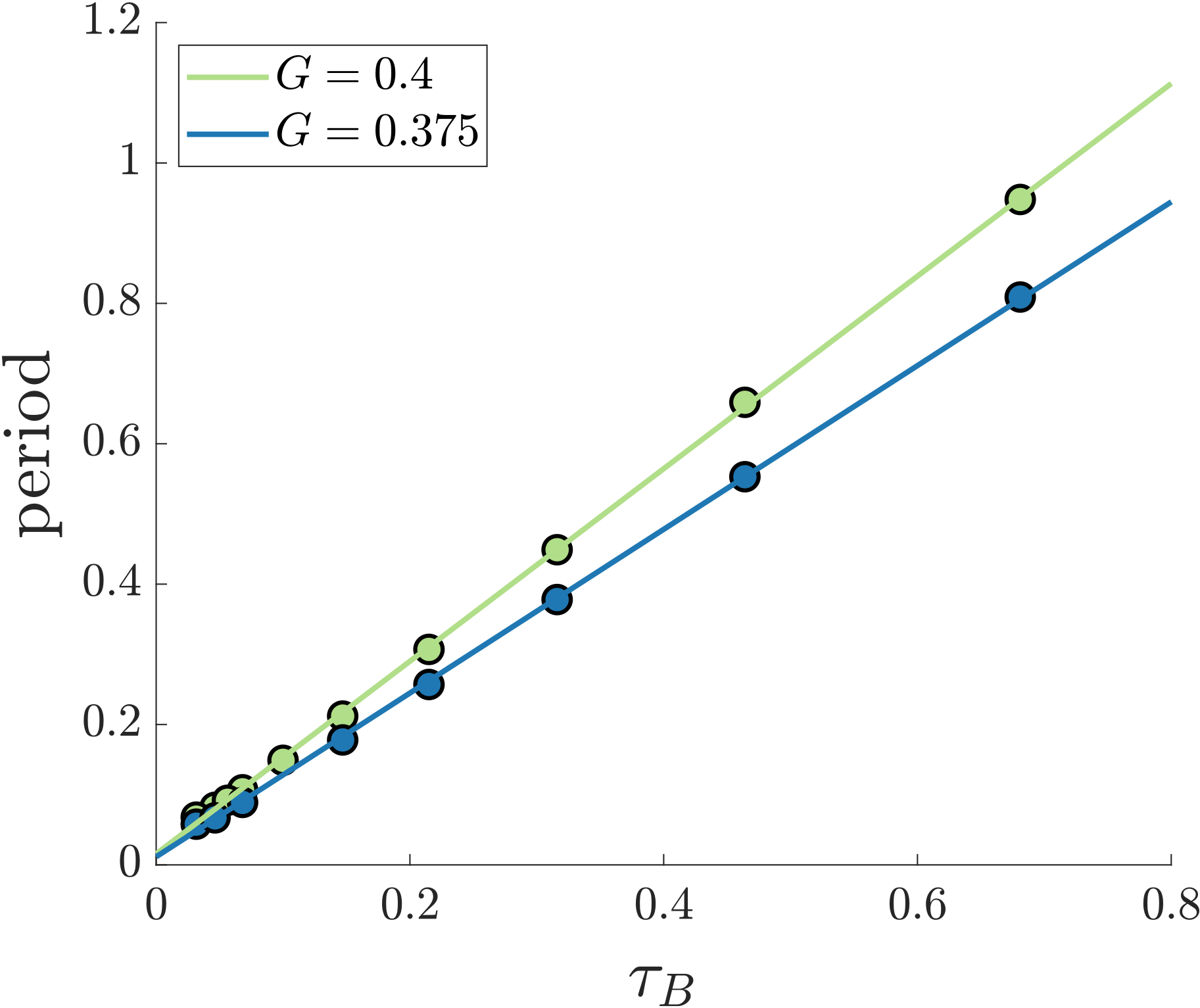}
        \caption{Period of the oscillations in amplitude in the nonlinear regime. $\tau_S = 0.001$, $G_B=G_S=G$.}
        \label{fig:nonLinPeriodTauB_G}
    \end{subfigure}
    \caption{\textbf{(a)-(b)} Phase diagrams of emergent consistent patterns in the nonlinear regime: coloured light blue for stationary (Fig.~\ref{fig:kymoStat}), dark blue for a travelling wave (Fig.~\ref{fig:kymoG02}), green for a standing wave (Fig.~\ref{fig:kymoG03}) and pink for a travelling wave with an oscillating amplitude (Fig.~\ref{fig:nonLinKymo}).  Each dot represents a simulation. They are classified using the maximal concentration difference on the domain and the location of the maximum of $c$ on the domain, more details are given in App.~\ref{app:simClassification}. The lines are analytically calculated boundaries, the dotted line for Eq.~\eqref{eq:mainCondition}, the dashed line for Eq.~\eqref{eq:b<0}, the dash-dot line for Eq.~\eqref{eq:c>0}. \textbf{(c)} Period of amplitude oscillations in the nonlinear regime. The lines are linear fits to the simulation results.}
    \label{fig:nonLin}
\end{figure}

To study the dynamics in the nonlinear regime in greater detail, we use a 1-dimensional model. We do this because the 1-dimensional model is much faster to solve and makes it easier to study the development of the solution over time. In App.~\ref{app:Lx_Ly}, it is shown that the 1-mode remains dominant when the domain is 1-dimensional. The nonlinear dynamics are studied by running simulations for various parameters. The solutions can be categorised in four groups: i) stationary solutions (Fig.~\ref{fig:kymoStat}), ii) standing waves (Fig.~\ref{fig:kymoG03}), iii) travelling waves (Fig.~\ref{fig:kymoG02}) and iv) travelling waves with an oscillating amplitude (Fig.~\ref{fig:nonLinKymo}). In \mbox{Figs. \ref{fig:kymoStat}-\subref{fig:nonLinKymo}}, two kymographs are shown for each solution.  Those on the left show the transition from the pattern in the linear regime to the developed nonlinear pattern. Those on the right show the developed dynamics of the nonlinear regime. In all cases, initially a standing wave pattern develops in the nonlinear regime. It develops even if there were no complex eigenvalues in the linear stability analysis. The standing wave pattern can then either persist (Fig.~\ref{fig:kymoG03}) or dissipate. If it dissipates one of the three other categories will develop. To determine when each type of solution occurs, we ran various simulations. The results are shown in Fig.~\ref{fig:nonLin}, the classification of the numerical solutions is described in App.~\ref{app:simClassification}. 
We found that if the difference in relaxation times is large and the solution is not close to being linearly stable, then a travelling wave with an oscillating amplitude develops. When the difference in relaxation times becomes smaller, the other three types of solutions occur: stationary solutions for larger elastic moduli and travelling or standing waves for smaller elastic moduli. There is one exception to this, the solution in Fig.~\ref{fig:nonLinClassTauB_G} at $G=0.45$ and $\tau_B = 10^{-0.5}$ is classified as a travelling wave instead of stationary, as was expected. This is because the velocity of the travelling wave did not decrease rapidly enough and a longer simulation would be needed for it to reach zero. To study the cause of the oscillation amplitude in case iv, we plot the period w.r.t. large relaxation time $\tau_B$ in Fig.~\ref{fig:nonLinPeriodTauB_G}. The results indicate that the period is linearly dependent on the long relaxation time. From this, we conclude that dissipation of the viscoelastic stress is an underlying mechanism for the periodic behaviour.

\section{Discussion}
The actin cytoskeleton is an active viscoelastic gel at the periphery of animal cells that regulates cell shape, tissue organisation and cell migration \cite{salb12, Bailles2019}. 
With the goal of enhancing our understanding of emergent pattern formation of the actin cytoskeleton in animal cells, we investigated here a minimal model of a flat active viscoelastic surface using the upper convected surface Maxwell model. The model contains a concentration field subject to diffusion and advection, that regulates the strength of the active stress. 
Our analysis discloses that the existence of two distinct viscoelastic relaxation time scales for shear and bulk elasticity is a mechanism that suffices to generate oscillations in this dynamical system.

We performed linear stability analysis predicting parameter regimes where the system is unstable and/or exhibits time-periodic oscillations. Corresponding results were confirmed with numerical simulations. We showed that for oscillations to emerge, two conditions have to be met; the elastic moduli have to be large enough w.r.t. the active term, $G_B + G_S > f'(1)$. Furthermore, the relaxation times have to be different from each other, $\tau_B \neq \tau_S$. 

Studying the nonlinear dynamics via simulations, we showed that four different types of solutions can occur. Depending on the parameters, we found stationary solutions, travelling waves, standing waves and travelling waves with an oscillating amplitude. Correspondingly, we disclosed that the minimal model under consideration can result in a variety of oscillatory dynamics which are also observed in cellular systems \cite{Bailles2019, Allard2013, mitsushima2010}.

Finally, we note that an important conclusion from our study is the hitherto unappreciated insight that viscoelastic mechano-chemical systems with two distinct relaxation times require only one concentration field to generate oscillations. In this respect, these systems differ fundamentally from pattern-forming reaction-diffusion systems \cite{cross_pattern_1993}.

\section*{ Acknowledgments} 
SA and EFF acknowledge support from the German Research Foundation DFG (grant AL1705/6 and FI 2260/5) from DFG Research Unit FOR-3013. 
EFF was further supported by the Heisenberg program – project number 495224622 (FI 2260/8-1) - and the Deutsche Forschungsgemeinschaft under Germany's Excellence Strategy, EXC-2068-390729961, Cluster of Excellence Physics of Life of TU Dresden. Simulations were performed at the Centre for Information Services and High Performance Computing (ZIH) at TU Dresden.

\section*{Appendix}
\begin{appendix}
\section{Linear stability analysis}
\label{app:linStabAna}
To derive the eigenvalues and their properties we continue the linear stability analysis starting from Eq.~\ref{eq:v_dot_kVE2}. Substituting this expression in Eq.~\eqref{eq:concentrationPerturb} and multiplying with the term in the denominator of Eq.~\eqref{eq:v_dot_kVE2}, we obtain
\begin{dmath}
\left(G_S\left(\lambda + \frac{1}{\tau_B}\right) + G_B \left(\lambda + \frac{1}{\tau_S}\right) \right)  \left( \lambda \delta c^{\vv{k}} + \vv{k}^2 \delta c^{\vv{k}} \right) =  \left(\lambda + \frac{1}{\tau_B}\right)\left(\lambda + \frac{1}{\tau_S}\right) \left( f'(1) \delta c^{\vv{k}} \right).
\end{dmath}
This can be rewritten to a parabolic equation w.r.t. $\lambda$, 
\begin{align}
    \begin{split}
        0 &= \lambda^2 \left( (G_S + G_B) - f'(1) \right) \delta c^{\vv{k}}\\
        &+ \lambda \left( \left( \frac{G_S}{\tau_B} + \frac{G_B}{\tau_S} \right) + (G_S + G_B)\vv{k}^2 - \left( \frac{1}{\tau_B} + \frac{1}{\tau_S} \right)  f'(1) \right) \delta c^{\vv{k}}\\
        &+ \left( \left( \frac{G_S}{\tau_B} + \frac{G_B}{\tau_S} \right) \vv{k}^2 - \frac{1}{\tau_B\tau_S} f'(1) \right) \delta c^{\vv{k}}.
    \end{split}
\end{align}
Dividing out the perturbation $\delta c^{\vv{k}}$ and multiplying with $\tau_S \tau_B$ gives,
\begin{align}
   \begin{split}
       0 &= \tau_B \tau_S \left(G_S + G_B - f'(1)\right) \lambda^2\\
       &+ \left(\tau_B G_B + \tau_S G_S + \tau_B \tau_S (G_B+G_S) \vv{k}^2 -(\tau_B+\tau_S)f'(1) \right)\lambda \\
       &+ (\tau_B G_B + \tau_S G_S) \vv{k}^2 -f'(1).
        \label{eq:lambda_quadratic}
   \end{split} 
\end{align}
For readability, we rewrite Eq.~\eqref{eq:lambda_quadratic} as $a\lambda^2 + b\lambda+c = 0$, the coefficients are given by Eqs.~\eqref{eq:a}, \eqref{eq:b} and \eqref{eq:c}.
The zeros of this quadratic equation are the eigenvalues. This potentially results in two eigenvalues $\lambda^+$ and $\lambda^-$, defined as $\lambda^{\pm} = \frac{-b \pm \sqrt{b^2 -4ac}}{2a}$ which in full is
\begin{dmath}
\scalebox{1.1}{$\lambda^\pm = -\frac{\tau_B G_B + \tau_S G_S + \tau_B \tau_S (G_B+G_S) \vv{k}^2 -(\tau_B+\tau_S)f'(1)}{2\tau_B \tau_S (G_S + G_B - f'(1))} \pm$}\\
\scalebox{1.1}{$ \frac{\sqrt{\left( \tau_B G_B + \tau_S G_S + \tau_B \tau_S (G_B+G_S) \vv{k}^2 -(\tau_B+\tau_S)f'(1) \right)^2 - 4\tau_B \tau_S (G_S + G_B - f'(1))\left( (\tau_B G_B + \tau_S G_S) \vv{k}^2 -f'(1) \right)}}{2\tau_B \tau_S (G_S + G_B - f'(1))}.$}
\label{eqApp:lambda}
\end{dmath}
Depending on the term in the root, i.e. the discriminant $\mathcal{D} = b^2 - 4ac$, the eigenvalues are complex. To describe the dependence on the viscous parameters, the eigenvalues can also be given in terms of the viscosities. For this use the relation $G_B = \frac{\eta_B}{\tau_B}$, idem for $G_S$. Substituting these into Eq.~\eqref{eq:v_dot_kVE1} results in the following expression,
\begin{equation}
    0 = -\eta_S \frac{\vv{k}^2 \delta \vv{v}^{\vv{k}}}{\tau_S \lambda +1} - \eta_B \frac{\vv{k} \cdot \delta \vv{v}^{\vv{k}} \vv{k}}{\tau_B \lambda +1} + i f'(1) \delta c^{\vv{k}} \vv{k}.
    \label{eq:stressIntegralViscosity}
\end{equation}
We can then do the same steps as before resulting in the following equation for the eigenvalues,
\begin{dmath}
\scalebox{1.2}{$\lambda^\pm = -\frac{\eta_B + \eta_S - f'(1)(\tau_B + \tau_S) + \vv{k}^2(\eta_B \tau_S + \eta_S \tau_B)}{2\left(\eta_B \tau_S + \eta_S \tau_B - f'(1)\tau_B\tau_S\right)}\pm $}\\
\scalebox{1.2}{$\frac{\sqrt{\left( \eta_B + \eta_S - f'(1)(\tau_B + \tau_S) + \vv{k}^2(\eta_B \tau_S + \eta_S \tau_B) \right)^2 - 4\left(\eta_B \tau_S + \eta_S \tau_B - f'(1)\tau_B\tau_S\right)\left(\vv{k}^2(\eta_B + \eta_S) - f'(1)\right)}}{2\left(\eta_B \tau_S + \eta_S \tau_B - f'(1)\tau_B\tau_S\right)}.$}
\label{eq:lambdaEta}
\end{dmath}
To study the eigenvalues in the viscous limit, take $\tau_S, \tau_B \rightarrow 0$ in Eq.~\eqref{eq:stressIntegralViscosity}, resulting in
\[
0 = -\eta_S \vv{k}^2 \delta \vv{v}^{\vv{k}} - \eta_B \vv{k} \cdot \delta \vv{v}^{\vv{k}} \vv{k} + i f'(1) \delta c^{\vv{k}} \vv{k}.
\]
Then, using the same steps as before gives the following equation for the eigenvalue in the viscous limit,
\begin{equation}
\lambda = \frac{f'(1)}{\eta_B + \eta_S} - \vv{k}^2.
\label{eq:viscousEigenvalues}
\end{equation}
So in the viscous limit, higher order modes are more stable and the 1-mode is the dominant mode as well.

\subsection{The special case of equal relaxation times}
\label{app:equalRelaxationTimes}
We found that we only get oscillatory behaviour if the relaxation times $\tau_S$ and $\tau_B$ are different. To show this, first consider equal relaxation times, i.e. $\tau_B = \tau_S = \tau$. The stability analysis starts the same as before, until Eq.~\eqref{eq:v_dot_kVE2}. Here we divide by $\lambda + \frac{1}{\tau}$ giving,
\begin{equation}
    \vv{k} \cdot \delta \vv{v}^{\vv{k}} = \frac{\lambda + \frac{1}{\tau}}{G_S + G_B} i f'(1) \delta c^{\vv{k}}.
    \label{eq:k_dot_deltaV}
\end{equation}
Substitute this in Eq.~\eqref{eq:concentrationPerturb} and multiply with $G_S+G_B$ to get
\begin{dmath}
\left(G_S + G_B \right)\left( \lambda \delta c^{\vv{k}} + \vv{k}^2 \delta c^{\vv{k}} \right) =  \left(\lambda + \frac{1}{\tau}\right) \left( f'(1) \delta c^{\vv{k}} \right).
\end{dmath}
From this equation, the following equation for $\lambda$,
\begin{equation}
    \lambda = \frac{f'(1)/\tau -\vv{k}^2(G_B + G_S)}{G_S + G_B - f'(1)},
\end{equation}
can be derived. So $\lambda$ is real and there will be no oscillatory behaviour in the linear regime.

\subsection{Existence of complex eigenvalues}
\label{app:complexEig}
To find the conditions for oscillations we consider the discriminant $\mathcal{D} = b^2 - 4ac$ of Eq.~\eqref{eq:lambda_quadratic}. There are complex eigenvalues if $\mathcal{D}<0$ for at least one $\vv{k}^2$. We can find a condition for this by considering that $\mathcal{D}$ is again a quadratic equation w.r.t. $\vv{k}^2$,  $\mathcal{D} = \tilde{a} \left(\vv{k}^2\right)^2 + \tilde{b} \vv{k}^2 + \tilde{c}$. The coefficients are defined as
\begin{align}
    \tilde{a} &= {\tau_B}^2 \,{\tau_S}^2 \,{{\left(G_B +G_S \right)}}^2, \label{eq:abeta}\\
    \begin{split}
    \tilde{b} &= 2\,{\left(G_B \,\tau_B \,\tau_S +G_S \,\tau_B \,\tau_S \right)}\,{\left(G_B \,\tau_B -f'(1) \,{\left(\tau_B +\tau_S \right)}+G_S \,\tau_S \right)} \\ &\qquad-4\,\tau_B \,\tau_S \,{\left(G_B \,\tau_B +G_S \,\tau_S \right)}\,{\left(G_B +G_S -f'(1) \right)}, \label{eq:bbeta} 
    \end{split}\\
    \tilde{c} &= {{\left(G_B \,\tau_B -f'(1) \,{\left(\tau_B +\tau_S \right)}+G_S \,\tau_S \right)}}^2 +4\,f'(1) \,\tau_B \,\tau_S \,{\left(G_B +G_S -f'(1) \right)}. \label{eq:cbeta}
\end{align}
$\mathcal{D}$ is a valley parabola since $\tilde{a} > 0$. So $\mathcal{D}(\vv{k})<0$ if and only if there exist $k^-, k^+ \in \mathbb{R}$ s.t. $\mathcal{D}(k^-) = \mathcal{D}(k^+) = 0$ and $k^- < \vv{k}^2 < k^+$. The values of $k^{\pm}$ are defined as $\frac{-\tilde{b} \pm \sqrt{\tilde{D}}}{2 \tilde{a}}$. Here $\tilde{\mathcal{D}}$ is again a discriminant and defined as $\tilde{\mathcal{D}} = \tilde{b}^2 -4\tilde{a}\tilde{c}$. For $k^-$ and $k^+$ to exist, $\tilde{\mathcal{D}}$ needs to be greater than 0. Substituting $\tilde{a}$, $\tilde{b}$ and $\tilde{c}$ results in
\begin{equation}
\tilde{\mathcal{D}} = 16\,G_B \,G_S \,f'(1) \,{\tau_B}^2 \,{\tau_S}^2 \,{{\left(\tau_B -\tau_S \right)}}^2 \,{\left(G_B +G_S -f'(1) \right)}.
\label{eq:Dbeta}
\end{equation}
Which is greater than 0 if and only if
\begin{equation}
    G_S + G_B - f'(1) > 0.
\end{equation}
So complex eigenvalues can only exist if $G_B + G_S > f'(1)$. Note that this only implies that complex eigenvalues can exist, but if there is no mode $\vv{k}$ such that $k^- < \vv{k}^2 < k^+$, then there are no complex eigenvalues.

\subsection{Stability}
\label{app:stability}
The system is stable if $\text{Re}(\lambda^\pm) < 0$ for all modes $\vv{k}$. We consider two cases $G_B+G_S$, smaller than $f'(1)$ and greater than $f'(1)$. If $G_B+G_S<f'(1)$ then $\lambda^{\pm}$ is real and the coefficient $a$ in Eq.~\eqref{eq:a} is negative. So the system is stable if 
\[
\lambda^- = \frac{-b - \sqrt{\mathcal{D}}}{2a} < 0.
\]
Because $a<0$ and $\mathcal{D} > 0$ it holds that $\lambda^- > \frac{-b}{2a}$. By definition  (Eq.~\eqref{eq:b}) $b$ grows with $\vv{k}^2$. Since $\lambda^- > \frac{-b}{2a}$, $\lambda^-$ grows when increasing $\vv{k}^2$ as well. This would result in arbitrarily large eigenvalues, hence linear stability analysis cannot be used in this case. 
To find the stability conditions for the second case we split the calculation into several steps. 

\subsubsection*{Proof of $\lambda^\pm_{|\vv{k}^2=0} \in \mathbb{R}$}
Before calculating the sign of $\text{Re}(\lambda^\pm)$ we first show that $\lambda^\pm_{|\vv{k}^2=0} \in \mathbb{R}$. This is equivalent to $\mathcal{D}_{|\vv{k}^2=0} \geq 0$, which follows directly from the definition of $\tilde{c}$ in Eq.~\eqref{eq:cbeta} and $\mathcal{D}_{|\vv{k}^2=0} = \tilde{c}$.

\subsubsection*{Proof of $\lambda^-(0) < 0 < \lambda^+(0)$}
To prove that $\text{Re}(\lambda^+) < 0$ for large $\vv{k}^2$ we first show that $\lambda^-(0) < 0 < \lambda^+(0)$. Since $a>0$, this is equivalent to $-b - \sqrt{D} < 0 < -b + \sqrt{\mathcal{D}}$. This holds if $b^2 < \mathcal{D}$ for $\vv{k}^2 = 0$, which follows from $c_{|\vv{k}^2 = 0} = -f'(1) < 0$.

\subsubsection*{Proof of $\lambda^\pm(k^+) < 0$}
Now we show that $\lambda^\pm(k^+) < 0$. The discriminant $\mathcal{D}$ at $\vv{k}^2 = k^+$ is zero by definition, so $\lambda^+(k^+) = \lambda^-(k^+)$. Because $a$ is positive if $G_B + G_S > f'(1)$ we only have to show that $b(k^+) > 0$. The definition of $k^+$ is $\frac{-\tilde{b} + \sqrt{\tilde{D}}}{2\tilde{a}}$, substituting $k^+$ in $b$ gives
\[
b(k^+) = \tau_B G_B + \tau_S G_S -(\tau_B + \tau_S)f'(1) + \tau_B \tau_S(G_B + G_S)\frac{-\tilde{b} + \sqrt{\tilde{D}}}{2\tilde{a}}.
\]
Because $G_S + G_B > f'(1)$ and $\tau_B \neq \tau_S$, we know that $\tilde{D} > 0$ and by definition we have that $\tilde{a}>0$. So to show that $b(k^+) > 0$ it suffices to show that
\[
0<2\tilde{a} \left( \tau_B G_B + \tau_S G_S -(\tau_B + \tau_S)f'(1) \right) - \tau_B \tau_S(G_B + G_S)\tilde{b}.
\]
Filling in $\tilde{a}$ and $\tilde{b}$ results in
\[
0< 3 \left(G^2_B\tau_B + G_S^2 \tau_S G_BG_S(\tau_B + \tau_S) - f'(1)(G_B \tau_B + G_S \tau_S)\right) + f'(1)(G_B\tau_S + G_S \tau_S).
\]
Using that $f'(1) < G_B + G_S$ it suffices to show that 
\[
0< 3 \left(G^2_B\tau_B + G_S^2 \tau_S G_BG_S(\tau_B + \tau_S) - (G_B+G_S)(G_B \tau_B + G_S \tau_S)\right) + f'(1)(G_B\tau_S + G_S \tau_S).
\]
Which is equivalent to 
\[
0<f'(1)(G_B \tau_S + G_S \tau_B).
\]

\subsubsection*{Proof of $\lambda^+(\vv{k}^2) < 0$ for $\vv{k}^2 > k^+$}
\begin{SCfigure}
    \centering
    \includegraphics[width=0.4\textwidth]{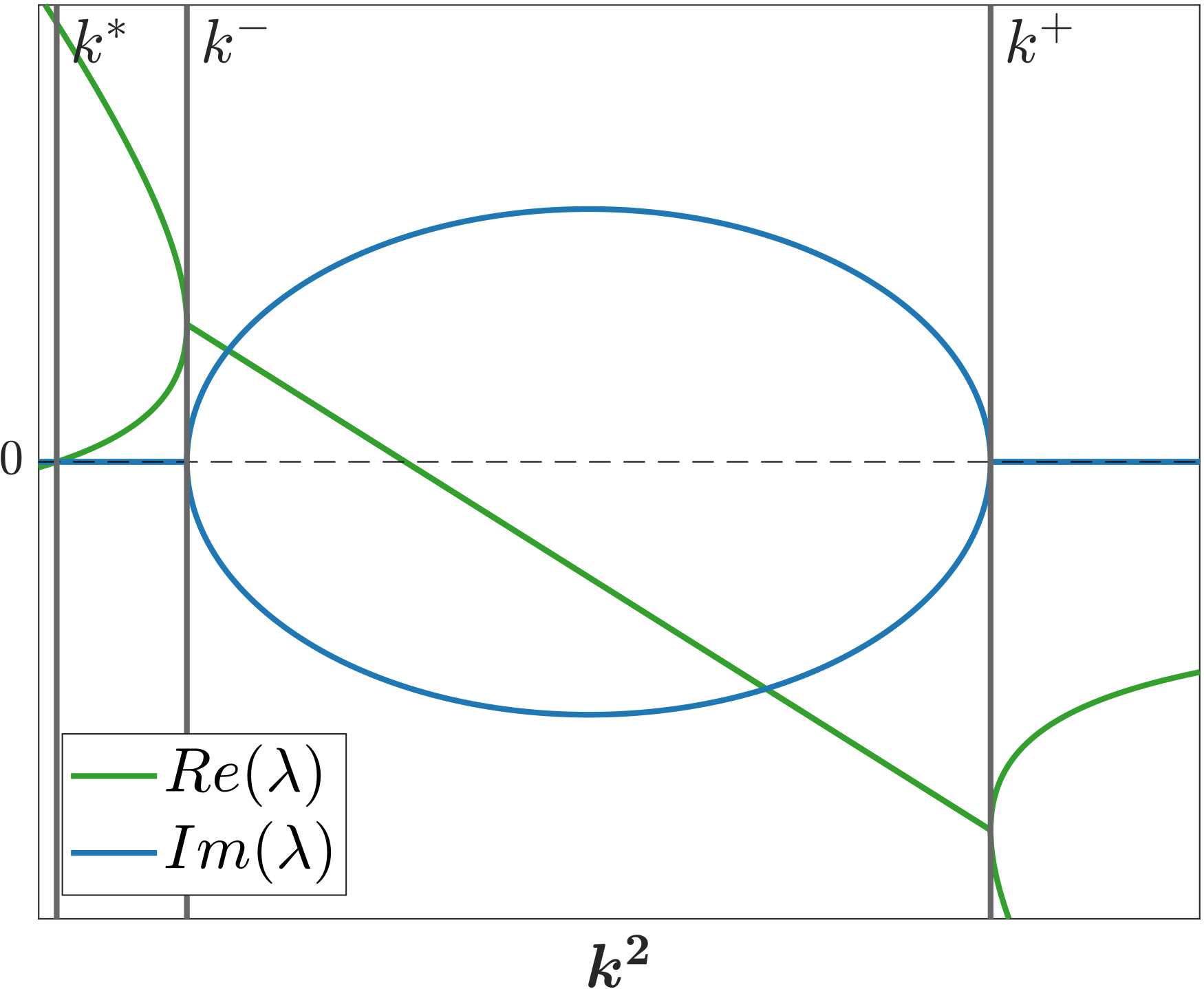}
    \caption{Sketch of the eigenvalues $\lambda^\pm$. The vertical lines denote the values of $k^*$, $k^-$ and $k^+$.}
    \label{fig:lambdaSketch}
\end{SCfigure}

To show this we first consider the zeros of $\lambda^\pm$ on the domain where $\mathcal{D} > 0$. By definition $\lambda^\pm = \frac{-b \pm \sqrt{b^2 - 4ac}}{2a}$, so the only zero is at $c=0$. We call this zero $k^*$, $k^* = \frac{f'(1)}{\eta_S + \eta_B}$ (see Fig.~\ref{fig:lambdaSketch}). So $\lambda^\pm$ has only one zero on the domain where $\mathcal{D} > 0$. We know that $\lambda^+(0) > 0 > \lambda^-(0)$ and $\lambda^+(k^-) = \lambda^-(k^-)$, so $k^* < k^-$. This in combination with $\lambda^+(k^+)<0$ implies that $\lambda(\vv{k}^2) < 0$ for $\vv{k}^2 > k^+$.

\subsubsection*{Finding the dominant mode}
The mode $\vv{k}$ with the largest $\text{Re}\left(\lambda^+(\vv{k}^2)\right)$ is the dominant mode. To find the dominant mode we first consider $\partial_{\vv{k}^2} b$, which is equal to $\tau_B \tau_S(G_B + G_S) >0$. This implies that when $\vv{k}^2 \in [k^-, k^+]$, the eigenvalue decreases for increasing $\vv{k}^2$. Now consider $\partial_{\vv{k}^2} \mathcal{D}$, we know that $\mathcal{D} = \tilde{a} (\vv{k}^2)^2 + \tilde{b}\vv{k}^2 + \tilde{c}$ and that $\tilde{a} > 0$. Thus $\partial_{\vv{k}^2} \mathcal{D} < 0$ for $\vv{k}^2 < \frac{-\tilde{b}}{2 \tilde{a}}$. $\frac{-\tilde{b}}{2\tilde{a}}$ is greater than $k^-$ by definition. So $\partial_{\vv{k}^2}\mathcal{D}$ is negative for $\vv{k}^2 < k^-$. This in combination with the fact that $\partial_{\vv{k}^2} b<0$ implies that $\partial_{\vv{k}^2}\text{Re}(\lambda^+)<0$ for $\vv{k}^2 < k^+$.
In the previous subsection we showed that $\text{Re}(\lambda^+(0))>0$ and $\text{Re}(\lambda^\pm) < 0$  for $\vv{k}^2> k^+$. In conclusion, the dominant eigenvalue is for the smallest $\vv{k}^2$. This would be the 0-mode ($\vv{k}^2 = 0$), however the corresponding coefficient $\delta c^{\vv{0}}$ is zero. Therefore, the dominant mode is the next smallest mode, which is the 1-mode $\vv{n}^2 = 1$ ($\vv{k}^2 = 4\pi^2$).

\subsubsection*{The amplitude of the 0-mode vanishes}
\label{app:0mode}
We argue that $\delta c^{\vv{0}}$ is equal to 0. If it were not this would result in an increase/decrease of the total concentration $\int_\Omega c d \Omega$. However, we can show that this is constant. To do so, integrate Eq.~\eqref{eq:c_scaled} over the entire domain,
\[
\int_\Omega \left( \partial_t c + \nabla \cdot (c \vv{v}) - \Delta c \right) d \Omega = 0.
\]
Split the integral into three separate integrals,
\[
\partial_t \int_\Omega c d \Omega + \int_\Omega \nabla (c \vv{v}) d \Omega + \int_\Omega \Delta c d \Omega = 0.
\]
Because of the divergence theorem, the second and third integral are equal to $\int_\Gamma c \vv{v} \cdot \vv{n} d \Gamma$ and $D_c \int_\Gamma \nabla c \cdot \vv{n} d\Gamma$, where $\Gamma$ is the boundary of $\Omega$. These integrals are zero because of the periodic boundary conditions. Therefore,  only $\partial_t \int_\Omega c d \Omega = 0$ remains.

\subsection{Rectangular domain}
\label{app:Lx_Ly}
For a rectangular domain,  the wave vector $\vv{k}$ of a mode is defined as $\vv{k} = 2 \pi \begin{pmatrix}  n_1 / L_x \\ n_2 /L_y\end{pmatrix}$, where $L_x$ and $L_y$ are the lengths of the sides of the domain in the $x$ and $y$ direction. $\vv{k}^2$ then becomes $4 \pi^2 \left((n_1/L_x)^2 + (n_2/L_y)^2\right)$. In the previous section we showed that the dominant modes are those for the lowest $\vv{k}$. Meaning that, if we assume that $L_x > L_y$ then the dominant mode will be for $\vv{n} = \begin{pmatrix} n_1 /L_x \\ 0\end{pmatrix}$. This shows that the longest wavelength is preferred. Additionally, if $L_y \ll 1$ then the norm of the nodes with $n_2 \neq 0$ will become very large. The eigenvalue decreases when increasing $\vv{k}^2$, so the dynamics will become 1-dimensional in the linear regime. Moreover, if we assume that the velocity in the $y$ direction is zero and all variables are constant w.r.t. $y$, then the full system of equations (Eqs.~\eqref{eq:devS_scaled}-\eqref{eq:force_scaled}) will also result in a 1-dimensional solution.

\section{Simulation}
\subsection{Implementation}
\label{app:implementation}
For the implementation we used the FEM C++ library Legacy AMDiS \cite{Vey2007, Witkowski2015}. We choose an IMEX time stepping  scheme, implementing all spatial derivatives implicitly. For example for the term $\nabla \vv{v} \bar{S}$ in Eq.~\eqref{eq:devS}, $\nabla \vv{v}$ is implemented implicitly and $\bar{S}$ explicitly. Doing so gave the most numerically stable simulations. The mesh is a triangulated $1\times 1$ square. For the viscoelastic stress we use the upper convected surface Maxwell model in Eqs.~\eqref{eq:devS} and \eqref{eq:trS}, and for the concentration $c$ we use Eq.~\eqref{eq:c_scaled}. To implement the force balance in Eq.~\eqref{eq:force_scaled} an additional artificial diffusion term, $D_{art} \Delta \vv{v}$, is required to dampen numerical oscillations. The choices for the numerical parameters are given in App.~\ref{app:verification}. The equation becomes
\begin{equation}
    \rho \partial_t^\bullet \vv{v} = \nabla \cdot \bar{S} + \nabla \left(\frac{1}{2} \tr (S) + f(c) \right) + D_{art} \Delta \vv{v}.
    \label{eq:forceNum}
\end{equation}
Each time iteration the entire system of equations is solved. For this the following finite element space is used,
\begin{equation}
    \Psi = \Big\{ \psi \in C(\Omega) \cap L^2(\Omega) \Big| \evalat[\big]{\psi}{k} \in P_1(k), k \in T_\Omega \Big\},
\end{equation}
where $T_\Omega$ is the triangulation of $\Omega$. The weak form of the fully discrete coupled system reads: Find $c^{n+1}, \tr (S)^{n+1} \in \Psi$, $\vv{v}^{n+1} \in \Psi^2$, $ \bar{S}^{n+1} \in \Psi^{2\times2}$ such that for all $\psi, \phi \in \Psi$, $\vv{\psi} \in \Psi^2$, $\Phi \in \Psi^{2\times2}$ the following holds,
\begin{align}
    0 &= \int_{\Omega} \left( \left( \frac{c^{n+1} - c^n}{\Delta t} - (\vv{v}^n \cdot \nabla) c^{n+1} \right) \psi + \hat{D}_c \nabla c^{n+1} \cdot \nabla \psi \right) d\Omega, \label{eq:concentrationWeak} \\
    \begin{split}
    0 &= \int_\Omega  \left( \rho\left(\frac{\vv{v}^{n+1} - \vv{v}^{n}}{\Delta t} + (\vv{v}^{n}\cdot \nabla) \vv{v}^{n+1} \right) - \nabla \cdot \bar{S}^{n+1} - \frac{1}{2} \tr (S)^{n+1} - f'(c^{n})\nabla c^{n+1} \right) \cdot \vv{\psi} d\Omega \\
    &+ \int_\Omega \left( D_{art} \left( \partial_x \vv{v}^{n+1} \cdot \partial_x \vv{\psi} + \partial_y \vv{v}^{n+1} \cdot \partial_y \vv{\psi} \right) \right) d\Omega,
    \end{split} \label{eq:forceBalanceWeak}\\
    0 &= \int_\Omega \left( \tr (S)^{n+1} + \tau_B\left( \frac{\tr (S)^{n+1} - \tr (S)^n}{\Delta t} - 2(\bar{S}^n:\nabla \vv{v}^{n+1}) + \tr (S)^n \nabla \cdot \vv{v}^{n+1}\right) - 2 \eta_B\nabla \cdot \vv{v}^{n+1} \right) \phi d \Omega, \label{eq:trSweak} \\
    \begin{split}
    0 &= \int_\Omega \left( \bar{S}^{n+1} - 2\eta_S \bar{D}^{n+1} \right) : \Phi d \Omega \\
    &+ \int_\Omega \tau_S \left( \frac{\bar{S}^{n+1} - \bar{S}^n}{\Delta t} + (\vv{v}^n\cdot\nabla) \bar{S}^{n+1} - \nabla \vv{v}^{n+1}\bar{S}^n - \bar{S}^n(\nabla \vv{v}^{n+1})^T + I (\bar{S}:\nabla \vv{v}^{n+1}) - \tr (S)^n \bar{D}^{n+1}\right)   : \Phi d \Omega.
    \end{split}
    \label{eq:SbarWeak}
\end{align}
The superscript $n$ denotes the time iteration.

\subsection{Verification and validation}
\label{app:verification}
\begin{figure}
    \centering
    \begin{subfigure}[t]{0.49\textwidth}
        \includegraphics[width = 0.99\textwidth]{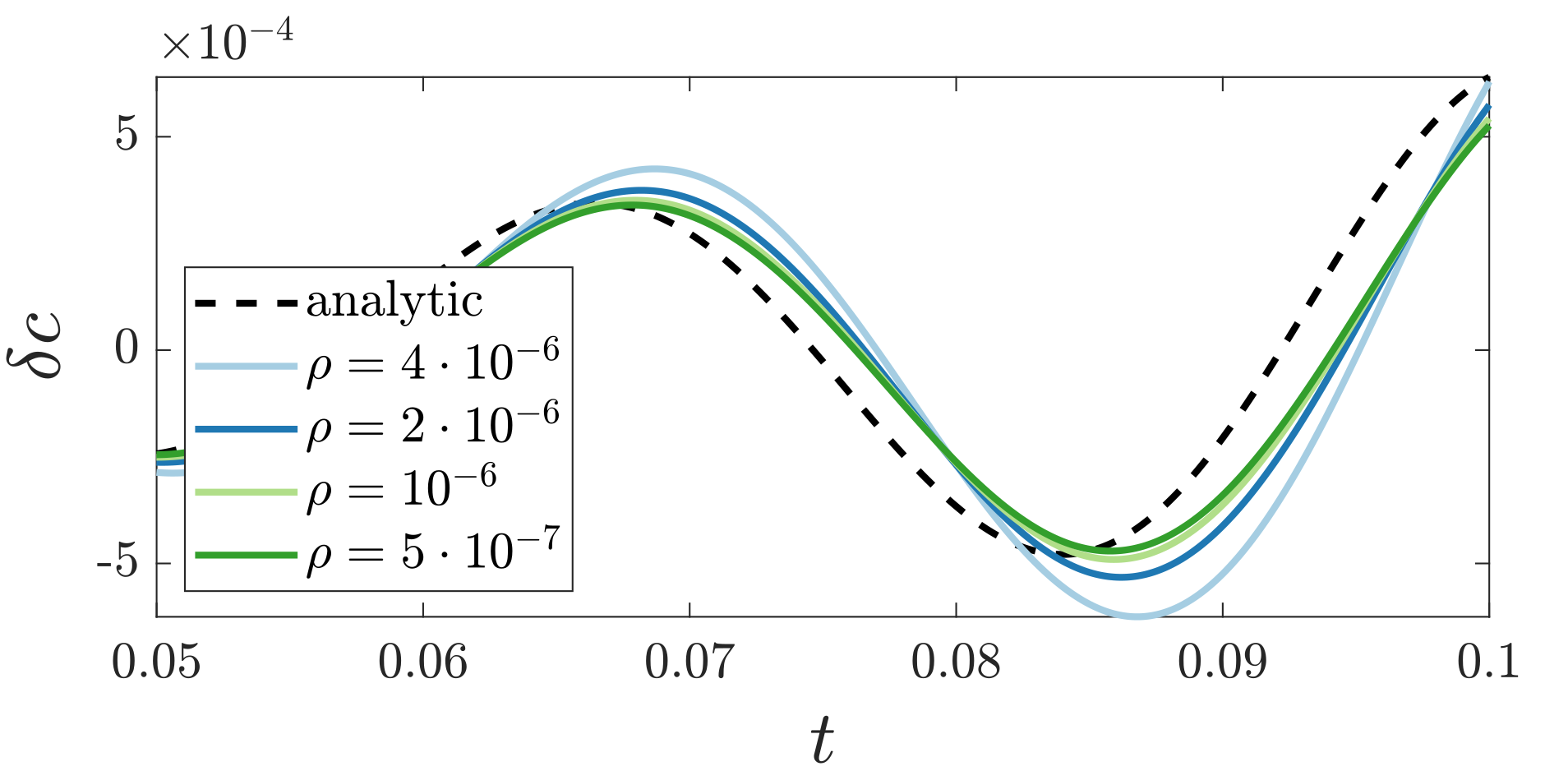}
        \captionsetup{width=0.8\textwidth}
        \caption{Comparison of numerical solutions with different densities to the analytic solution. $\tau_B = 0.1$, $\tau_S = 0.001$, $G_B = G_S = 0.45$.}
        \label{fig:rhoConv}
    \end{subfigure}
    \begin{subfigure}[t]{0.49\textwidth}
        \includegraphics[width = 0.99\textwidth]{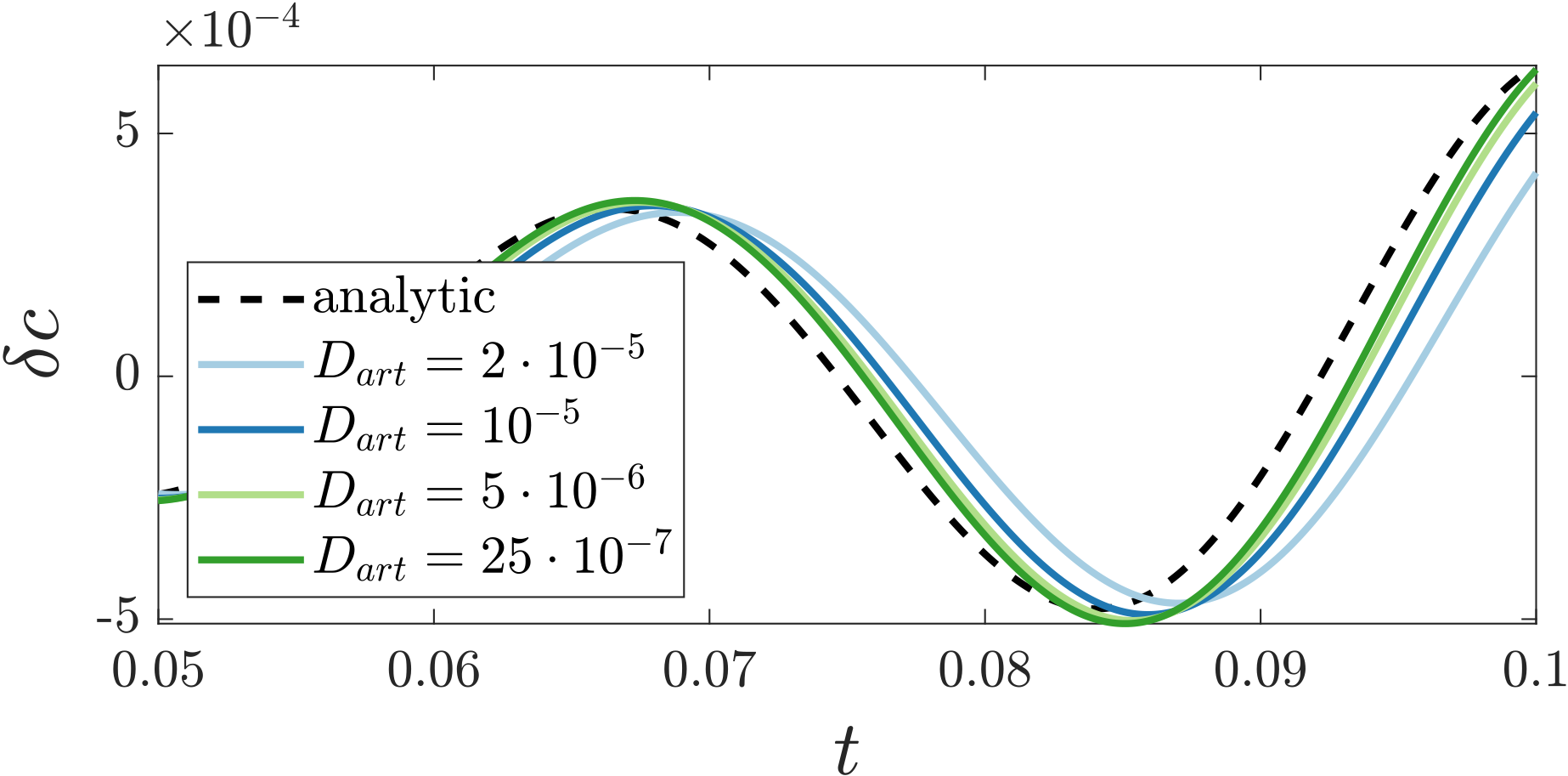}
        \captionsetup{width=0.8\textwidth}
        \caption{Comparison of numerical solutions with different artificial diffusion coefficients to the analytic solution. $\tau_B = 0.1$, $\tau_S = 0.001$, $G_B = G_S = 0.45$.}
        \label{fig:DartConv}
    \end{subfigure}\\
    \begin{subfigure}[t]{0.49\textwidth}
        \includegraphics[width = 0.99\textwidth]{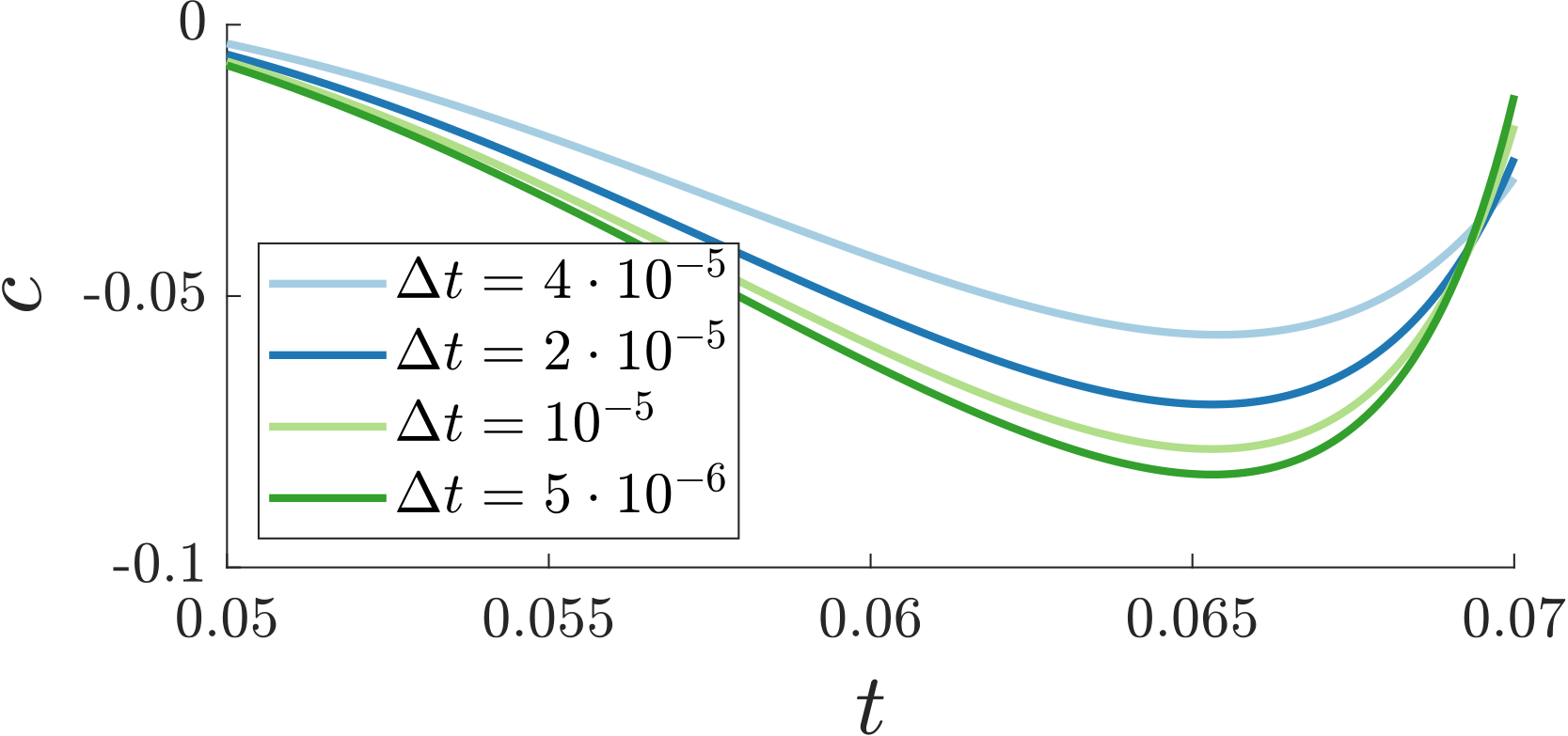}
        \captionsetup{width=0.8\textwidth}
        \caption{Convergence w.r.t. the time step size. $\tau_B = 0.1$, $\tau_S = 0.001$, $G_B = G_S = 0.4$.}
        \label{fig:timeStepConv}
    \end{subfigure}
    \begin{subfigure}[t]{0.49\textwidth}
        \includegraphics[width = 0.99\textwidth]{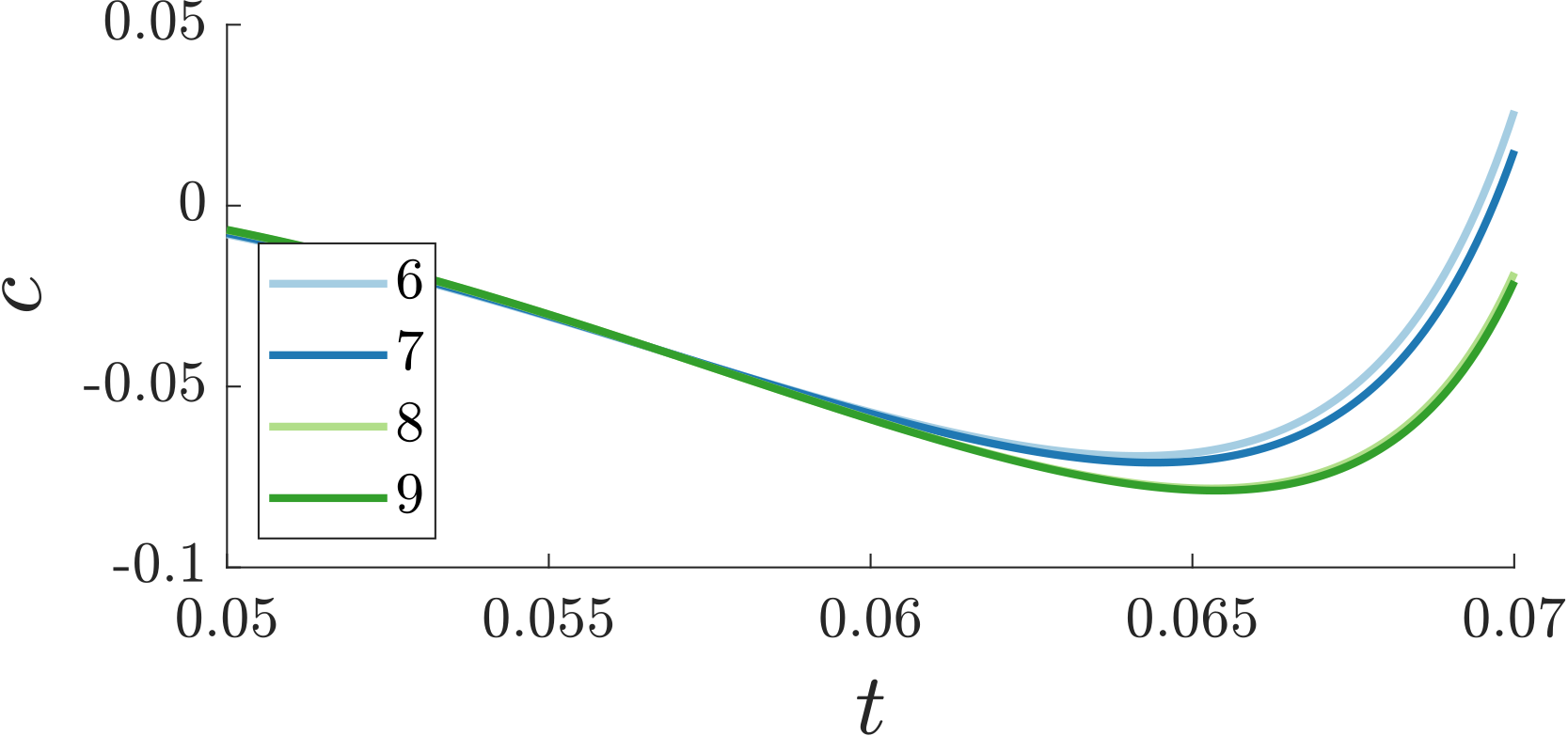}
        \captionsetup{width=0.8\textwidth}
        \caption{Convergence w.r.t. grid size, the numbers indicate bisections of the triangular grid. $\tau_B = 0.1$, $\tau_S = 0.001$, $G_B = G_S = 0.4$.}
        \label{fig:gridConv}
    \end{subfigure}
    \caption{Convergence of the numerical solution for the artificial diffusion $D_{art}$, density $\rho$ and the numerical parameters. $\delta c$ is the perturbation as defined in Eq.~\eqref{eq:c_perturb}. The figures show the solution at $x = 0.125$.}
    \label{fig:convergence}
\end{figure}

To validate the simulation we compare a specific case ($\tau_B = 0.1, \tau_S = 0.001, G_B = G_S = 0.45$) to the solution of the linear stability analysis. This case is chosen because $\text{Re}(\lambda^+)$ is close to zero, such that the solution remains close to the equilibrium value. As long as the concentration remains close to the equilibrium value, the evolution of all quantities will remain close to the definitions in Eqs.~\eqref{eq:c_perturb}-\eqref{eq:Sbar_perturb}. As initial condition, we choose $\delta c = \text{Re}(e^{2\pi i \vv{x}})$. Then, using the relations in Eqs.~\eqref{eq:concentrationPerturb}, \eqref{eq:trS_eq_perturb} and \eqref{eq:Sbar_eq_perturb}, we calculate $\delta \vv{v}$, $\delta \tr (S)$ and $\delta \bar{S}$ to use these as initial condition. 
To compare the numerical solution to the analytic solution we choose a point in the domain and compare the evolution of $\delta c$ at this point. 
Fig.~\ref{fig:rhoConv} shows that for decreasing density $\rho$ the amplitude of the numerical solution converges to the amplitude of the analytic solution (which was derived for $\rho=0$).
We further find that the numerical solution has a small delay w.r.t. the analytic one. This delay decreases when reducing the artificial diffusion $D_{art}$ (Fig.~\ref{fig:DartConv}).
This indicates that reducing $\rho$ and $D_{art}$ together gives convergent behaviour very close to the analytical solution. 
Note however, that these parameters cannot be reduced to zero in the numerical model as this would create numerical oscillations ($D_{art}=0$) or an unsolvable system ($\rho=0$).  

To investigate convergence in the numerical parameters, we consider a different test case $\tau_B = 0.1$, $\tau_S = 0.001$, $G_B = G_S = 0.4$) with a larger $\text{Re}(\lambda^+)$ because this resulted in a faster growth in amplitude. This faster growth made the simulation less numerically stable. Hence, numerical parameters that resulted in a stable simulation for this case, will also do so for more linear stable cases.
Reducing the time step or grid size results in convergence towards a solution close to the analytic solution. However, there remains a short delay in the numerical solution, probably caused by the artificial diffusion.
We also used the results of the second case to determine the values for the numerical parameters. We choose the artificial diffusion to be the smallest value that still suppresses the numerical oscillations. The density is chosen as the smallest value for which the simulation converges. For the numerical parameters we choose the coarsest value for which reducing it did not result in a significant improvement. We did this for both the 1-dimensional and the 2-dimensional implementation. Additionally, we compared the 1-dimensional solution to a 2-dimensional solution which was stationary w.r.t. $y$, to ensure these also resulted in the same solution

\subsection{Classification of the simulation results}
\label{app:simClassification}
For the classification of the simulations we looked at several values. The first is the maximal difference in concentration on the domain,
\begin{equation}
    \delta c_{max} = \max_{\vv{x} \in \Omega} c - \min_{\vv{x} \in \Omega} c.
    \label{eq:deltaCmax}
\end{equation}
This can also be interpreted as the amplitude of the oscillations. The second value is the correlation with the modes expressed by the correlation coefficient $\rho_{\vv{k}}$ for mode $\vv{k}$,
\begin{equation}
    \rho_{\vv{k}} =  \frac{\int_\Omega (c - c_{avg}) g_{\vv{k}} d\Omega}{\sqrt{\int_\Omega (c - c_{avg})^2 d\Omega \int_\Omega g_{\vv{k}}^2 d\Omega}},
\end{equation}
where $c_{avg} = \int_\Omega c d\Omega$ and $g_{\vv{k}} = \sin \left(2 \pi \vv{k} \cdot \vv{x} + \vv{b} \right)$. The point $\vv{b}$ is chosen between the minimum and maximum of $c$. The third value is the location of the maximum in concentration, $\vv{x}_{max} = \arg\max_{\vv{x} \in \Omega} c$. To classify the numerical solutions in the linear regime we used the following.
\begin{itemize}
    \item A numerical solution is categorised as unstable if 
    \[\delta c_{max}(t=0) < \delta c_{max}(t=T) \qquad \text{or} \qquad \max_{t\in[0.T]} \delta c_{max}(t) > 10 \delta c_{max}(t=0) .\] Here $T$ is the time the simulation finished.
    \item A numerical solution is categorised to have an oscillating $\vv{k}$-mode if \[\max_{t\in [0, T_1]} \rho_{\vv{k}}(t) > 0.8 \qquad \text{and} \qquad \min_{t\in [0,T_1]} \rho_{\vv{k}}(t) < -0.8.\]
    Here $[0, T_1]$ is a time interval where the numerical solution is still in the linear regime. $T_1$ is chosen as the last time $t$ where $|c - c_{avg}| < 0.2$ still holds.
\end{itemize}
To classify the simulations in the nonlinear regime we first calculate the time domain for which the solution is classified as nonlinear. For this we chose the interval $[T_{nl}, T]$, where $T_{nl}$ is the first time that $\delta c_{max} > 0.2$. If $T_{nl} < T-3$ then this simulation is not classified because the consistent patterns described in Sec.~\ref{sec:nonlin} might not have developed yet. Otherwise the nonlinear time domain is chosen as $[T-3, T]$. To classify the simulations we use the amplitude of $\delta c_{max}$ and the total distance travelled by the maximum of $c$, called $\alpha$ and $s_{max}$ respectively. The amplitude is defined as
\begin{equation}
    \alpha = \max_{t \in [T-3, T]} \delta c_{max} - \min_{t \in [T-3, T]} \delta c_{max}.
\end{equation}
The total distance travelled by the maximum is defined as $s_{max} = x_{max}(T) - x_{max}(T-3)$. The simulations are then classified in the four categories as follows;
\begin{itemize}
    \item[i] Stationary solution if $\alpha < 10^{-3}$ and $s_{max} \leq 0.03 $.
    \item[ii] Travelling wave if $\alpha < 10^{-3}$ and $s_{max} > 0.03 $. $s_{max}$ is chosen small here because these waves travelled much slower than those in category iv.
    \item[iii] Standing wave if $\alpha \geq 10^{-3}$ and $s_{max} < 0.5$.
    \item[iv] Travelling wave with oscillating solution if $\alpha \geq 10^{-3}$ and $s_{max} > 0.5$. 
\end{itemize}

\end{appendix}

\printbibliography

\end{document}